\documentclass[ reprint, amssymb, amsmath, aip, jcp]{revtex4-1}

\usepackage{comment,bbold,dcolumn,csquotes,xspace,graphicx,amsmath,tikz,multirow,bbold}
\usepackage[nolist]{acronym}
\usepackage[normalem]{ulem}
\usepackage[utf8]{inputenc}

\usepackage[colorlinks=true,linkcolor=blue]{hyperref}%
\expandafter\ifx\csname package@font\endcsname\relax\else
 \expandafter\expandafter
 \expandafter\usepackage
 \expandafter\expandafter
 \expandafter{\csname package@font\endcsname}%
\fi
\usetikzlibrary{positioning}

\newcommand{\manu}[1]{{\color{black} #1}}
\newcommand{\manulast}[1]{{\color{black} #1}}
\newcommand{\gj}[1]{{\color{black} #1}}

\newcommand{\ket}[1]{{\ensuremath{|#1\rangle}\xspace}}
\newcommand{\bra}[1]{{\ensuremath{\langle #1|}\xspace}}

\newcommand{\elemm}[3]{\bra{#1} #2 \ket{#3}}
\newcommand{\basisq}[0]{{\mathcal B}(\boldqb)}
\newcommand{\bigx}[0]{{\mathbf X}}
\newcommand{\intbigx}[0]{\int \text{d}\bigx}
\newcommand{\mueqx}[0]{\mueq(\bigx)}
\newcommand{\mux}[0]{\mu(\bigx)}

\newcommand{\mueq}[0]{\mu_{\text{eq}}}

\newcommand{\pb}[0]{{P}}

\newcommand{\qs}[0]{{q}} 
\newcommand{\qb}[0]{{Q}}

\newcommand{\ecano}[0]{F}
\newcommand{\ecanotot}[0]{\ecano_0^{\text{tot}}}

\newcommand{\fulh}[0]{\hat{\bf H}_{\text{tot}}}
\newcommand{\hathqm}[0]{\hat{\bf h}_{\qm}}

\newcommand{\hatboldts}[0]{\hat{\bf t}}
\newcommand{\hatboldvextqm}[0]{\hat{{\bf v}}^{\text{ext}}_{\qm}}
\newcommand{\boldvextqm}[0]{{\bf v}^{\text{ext}}_{\qm}}
\newcommand{\vneqm}[0]{v_{\text{ne}}}
\newcommand{\vextqm}[0]{v^{\text{ext}}_{\qm}}

\newcommand{\hatboldwqm}[0]{\hat{\bf w}_{\qm}}

\newcommand{\hatboldwqmmm}[0]{\hat{\bf W}_{\qm}^{\mm}}

\newcommand{\hatps}[1]{\hat{p}_{#1}}

\newcommand{\wqm}[0]{{w}_{\qm}}
\newcommand{\wqmmm}[0]{{W}_{\qm}^{\mm}}

\newcommand{\boldqs}[0]{{\bf \qs}}
\newcommand{\boldqps}[0]{{\bf \qs}'}

\newcommand{\boldwqmmm}[0]{{\bf W}_{\qm}^{\mm}}

\newcommand{\hathtotmm}[0]{\hat{{\bf V}}^{\text{tot}}_\mm}

\newcommand{\hathtotqm}[0]{\hat{{\bf h}}^{\text{tot}}_\qm}
\newcommand{\vmm}[0]{{{\bf {V}}}_\mm}

\newcommand{\varpb}[1]{{\pb}_{#1}}

\newcommand{\hmm}[0]{{\bf H}_{\mm}}
\newcommand{\boldtb}[0]{{\bf T}}

\newcommand{\wmm}[0]{{W}_{\mm}}
\newcommand{\boldwmm}[0]{{\bf W}_{\mm}}
\newcommand{\vextmm}[0]{V^{\text{ext}}_{\mm}}
\newcommand{\boldvextmm}[0]{{\bf V}^{\text{ext}}_{\mm}}
\newcommand{\varqb}[1]{{\qb}_{#1}}
\newcommand{\varqs}[1]{{\qs}_{#1}}

\newcommand{\boldqb}[0]{{\bf \qb}}

\newcommand{\boldpb}[0]{{\bf P}}

\newcommand{\eqrdm}{\mathcal{P}_{\text{eq}}}

\newcommand{\hateqrdm}{\hat{\mathcal{P}}_{\text{eq}}}
\newcommand{\eqrdmqm}{\rdmqm_{\text{eq}}}
\newcommand{\hateqrdmqm}{\hatrdmqm_{\text{eq}}}
\newcommand{\eqrdmmm}{\rdmmm_{\text{eq}}}
\newcommand{\eqz}[0]{Z}
\newcommand{\eqztot}[0]{\eqz^{\text{tot}}}
\newcommand{\eqzqm}[0]{\eqz^{\qm}}

\newcommand{\dqss}{\text{d}\qs}
\newcommand{\dqs}{\text{d}\boldqs}
\newcommand{\dqps}{\text{d}\boldqps}
\newcommand{\dqb}{\text{d}\boldqb}
\newcommand{\dpb}{\text{d}\boldpb}

\newcommand{\intdqss}{\int\dqss}
\newcommand{\intdqs}{\int\dqs}
\newcommand{\intdqps}{\int\dqps}
\newcommand{\intdqb}{\int\dqb}

\newcommand{\tr}{\text{Tr}}

\newcommand{\nmm}{N_{\mm}}
\newcommand{\nqm}{N_{\qm}}
\newcommand{\mm}{{\text{mm}}}
\newcommand{\mmb}{{\text{MM}}}
\newcommand{\qm}{{\text{qm}}}
\newcommand{\qmb}{{\text{QM}}}

\newcommand{\potcano}{\ecano}
\newcommand{\potcanoz}{\potcano_0}
\newcommand{\potcanotot}{\ecanotot}
\newcommand{\uint}{\mathcal{U}}

\newcommand{\entrop}{\mathcal{S}}

\newcommand{\fmm}{\mathcal{F}_\mm}
\newcommand{\fqm}{\mathcal{F}_\qm}

\newcommand{\hatrdmqm}{\hat{\rdmqm\,}}

\newcommand{\rdmqm}{{\boldsymbol{\rho}}}

\newcommand{\rdmmm}{\boldsymbol{p}}

\newcommand{\densmm}{n}
\newcommand{\densqm}{\rho}

\newcommand\mydots{\hbox to 0.9em{.\hss.\hss.}}

\newcommand{\kb}{}

\begin{document}

\title{A variational formulation of the free energy of mixed quantum-classical systems: coupling classical and electronic density functional theories}

\author{Guillaume Jeanmairet}
\email{guillaume.jeanmairet@sorbonne-universite.fr}
\affiliation{Sorbonne Université, CNRS, Physico-Chimie des Électrolytes et Nanosystèmes
Interfaciaux, PHENIX, F-75005 Paris, France}
\affiliation{Réseau sur le Stockage Électrochimique de l'Énergie (RS2E), FR CNRS
3459, 80039 Amiens Cedex, France}
\author{Maxime Labat}
\affiliation{Sorbonne Université, CNRS, Physico-Chimie des Électrolytes et Nanosystèmes
Interfaciaux, PHENIX, F-75005 Paris, France}
\author{Emmanuel Giner$^*$}%
\email{emmanuel.giner@lct.jussieu.fr}
\affiliation{Laboratoire de Chimie Théorique, Sorbonne Université and CNRS, F-75005 Paris, France}

\begin{abstract}
Combining classical density functional theory (cDFT) with quantum mechanics (QM) methods offers a computationally efficient alternative to 
traditional QM/molecular mechanics (MM) approaches for modeling mixed quantum-classical systems at finite temperatures. 
However, both QM/MM and QM/cDFT rely on {somewhat ambiguous} approximations\gj{, the two major ones being:
i) the definition of the QM and MM regions as well as the description of their coupling, and ii) the choice of the  methods and levels of approximation made to describe each region}. \\
This paper addresses the second point and develop an exact theoretical framework that allows us to
clarify the approximations involved in the QM/cDFT formulation. 
We therefore establish a comprehensive density functional theory (DFT) framework for mixed quantum-classical  
systems within the canonical ensemble. 
We start by recalling the expression of the adiabatic equilibrium density matrix for a mixed system made of $\nqm$ quantum and $\nmm$ classical particles, \manulast{which can be related to a partial Wigner transformation}. 
Then, we propose a variational formulation of the Helmholtz free energy in terms of the full\gj{, non-equilibrium, } QM/MM density matrix. 
Taking advantage of permutational symmetry and thanks to constrained-search methods, 
we reformulate the computation of the Helmholtz free energy using only the quantum and classical one-body densities. 
\gj{Therefore, this paper generalizes both cDFT and electronic DFT (eDFT) to QM/MM systems.} \\
\gj{We then reformulate the functional to make the standard eDFT and cDFT Levy–Lieb functionals explicitly appear, together with a new universal correlation functional for QM/MM systems. }
A mean-field approximation is finally introduced in the context of solvation problems and we discuss its connection with several existing
mixed cDFT-eDFT schemes. \manu{An extension to the semi-grand canonical ensemble, \gj{where the number of classical particles is allowed to fluctuate,} is provided in the supplementary materials.}
\end{abstract}

\maketitle

\section{Introduction}
The Hohenberg-Kohn  theorems\cite{HohKoh-PR-64} dramatically reduce the complexity of simulating a quantum ($\qmb$) molecular system by enabling a focus 
on the one-electron density instead of the full $N$-electron wave function. Central to this simplification is the unknown universal density functional, 
which was later given a more rigorous mathematical foundation by Levy and Lieb\cite{Levy-PNAS-79,Lieb-IJQC-83}. 
Through the self-consistent framework introduced by Kohn and Sham\cite{KohSha-PR-65}, along with extensive efforts to develop approximations 
to the exact exchange-correlation functional (see, for example, a recent review in Ref. \onlinecite{Toulouse-DFT-23}), electronic density functional theory (eDFT) 
has become the workhorse of quantum chemistry.

Although electronic structure calculations at zero temperature can routinely handle systems with hundreds to even thousands of atoms, 
accounting for finite-temperature effects becomes crucial in many applications. Such calculations must therefore be set in a statistical ensemble, 
where the leading-order entropic contribution arises from nuclear motion \manu{due to the typically large gaps in electronic energies}. 
In this context, \manu{because of the large mass of nuclei with respect to that of electrons, it is physically relevant} 
to maintain electrons at the quantum level and to treat nuclei as classical particles. 
Nevertheless, the forces acting on the nuclei are computed from the solutions of the quantum electronic ground state,  
\textit{i.e.} from the solutions of the Born Oppenheimer Hamiltonian. 
This forms the basis of \textit{ab-initio} molecular dynamics \cite{iftimie_ab_2005,silpar-prl-99} 
which recovers the classical nuclear entropic contributions through either classical Monte Carlo algorithms or 
classical trajectories under the ergotic hypothesis. 

\manu{
Nevertheless, it is important to keep in mind that nuclear quantum effects play a significant role for light nuclei and can substantially 
alter the rates of certain chemical processes, such as proton transfer 
reactions\cite{BalGroLesMcC-JPC-96,HanKap-JCP-05,KimHam-JCP-06,rev-cpet-10-a,rev-cpet-schiffer-10}. 
Moreover, such proton transfers can also be strongly influenced by the solvation effects, 
specially when polar solvents are used\cite{Kapral-JPCM-15}. 
In this context, two fundamentally different yet interconnected nuclear quantum effects are at work:
(i) the intrinsic delocalization of nuclear wave packets due to the quantum nature of nuclei, and
(ii) the nonadiabatic couplings that arise between various adiabatic electronic states.
Nuclear delocalization can be captured through explicit nuclear wave functions (see, for example, Ref. \onlinecite{Beck-mctdh-rev-00} 
for a review of time-dependent mean-field approaches, and Ref. \onlinecite{Hammes-JCP-21} for the nuclear-electron orbital approach) 
or via path-integral formalisms such as ring-polymer molecular dynamics\cite{rpmd-rev-13}.
Nonadiabatic couplings, on the other hand, are commonly addressed using surface hopping methods based 
on classical nuclear trajectories\cite{Tully-JCP-90,HamTul-JCP-94,BenTod-JCP-00,LiTulSchFri-JCP-05,RicTho-JCP-13,Ananth-JCP-13},
modified time-dependent mean-field techniques\cite{rev-mctdh-nac-08}, 
semiclassical expansions of the density matrix and its Wigner representation\cite{rev-mctdh-nac-08}, 
or even hydrodynamic formulations of quantum mechanics à la Bohm-de Broglie\cite{BurCed-JCP-01-A,BurCed-JCP-01-B,BurBag-CP-06}. 
In contrast, the present work adopts a fully classical treatment of the solvent molecules and focuses 
on clarifying the formally exact framework for a given QM/MM partitioning, as typically addressed in \textit{ab initio} molecular dynamics.
}

Despite its considerable achievements, \textit{ab-initio} molecular dynamics is computationally demanding due to the extensive number of 
$\qmb$ calculations required to sample the exponentially large phase space. 
The convergence rate of the statistics can also vary significantly depending on the targeted properties and conditions\cite{PanSpaHarSceGal-PNAS-13,RozPanGibGal-PNAS-18}. 
To alleviate \textit{ab-initio}  {molecular dynamics} computational burden, 
$\qmb$/$\mmb$\cite{WarLev-JMB-76,MulLynKar-JACS-00,SenThi-Ang-09,Groenhof-BS-13} methods were developed, 
where only a small, critical portion of the system is treated at the $\qmb$ level, while the remainder is handled at the classical level (MM). 
The interaction between the classical particles are modeled using parametrized force fields and their dynamics obey Newton's law of motion. 
\manu{The choice of the classical model to represent molecular entities together with the choice of the $\qmb$/$\mmb$ interaction lacks 
of a unique definition and consists into a whole field of research\cite{ff-rev-18_2,ff-rev-18,ff-ml-rev-21,ff-ml-rev-23,qmmm-rev-21,qmmm-rev-25} 
involving both physically-motivated and machine-learning oriented approaches. }
\manu{However, despite the significant gain in computational cost with respect to \textit{ab initio} molecular dynamics},  
sampling the phase space of large systems remains computationally intensive even in $\qmb$/$\mmb$ approaches. 

Inspired by the extension of eDFT to $\qmb$ systems at finite temperatures\cite{Mermin-PR-65}, density-based formalisms have emerged for describing purely classical objects in statistical mechanics, particularly in the form of classical density functional theory (cDFT)\cite{Evans-AP-79}. 
Originally rooted in liquid state theory, cDFT is essentially a classical counterpart to eDFT, reformulating the classical statistical mechanics problem of large ensembles 
of identical particles (atoms or molecules) in terms of one-body density functionals. cDFT, which is naturally formulated within the 
grand canonical ensemble, establishes the existence of a unique functional of the classical particle density that reaches its minimum at the equilibrium density. At this minimum, the functional is equal to the grand potential, which serves as the  thermodynamic potential of the grand canonical ensemble. Therefore, most equilibrium properties can be obtained through functional minimization over the classical one-body density rather than through tedious sampling of configurations in an exponentially large phase space. 

The efficiency of both eDFT and cDFT stems from a common principle: each method replaces the challenge of computing a high-dimensional object, such as the wavefunction in QM or the probability distribution in MM, with the optimization of a functional that depends only on a low-dimensional quantity, \textit{i.e.} the one-body density of either quantum electrons or classical molecules. The mathematical framework allowing for such a simplification is  the constrained-search principle introduced by Levy and Lieb\cite{Levy-PNAS-79,Lieb-IJQC-83} which enables, under certain constraints of representativity, to give an explicit expression of the universal functionals of eDFT and cDFT\cite{DwaSch-PRE-11}, although not usable in practical calculations.  

{Its computational efficiency makes cDFT appealing}  to model solvent effect at the classical level while retaining a quantum description of the solute\cite{tang_solvent_2020,petrosyan_joint_2005,petrosyan_joint_2007,JeaLevBor-JCTC-20,LabGinJea-JCP-24}. In this approach, the total Hamiltonian is partitioned into well defined QM and MM Hamiltonians. 
However, just as in molecular dynamics based QM/MM approaches, the coupling interaction between the QM and MM regions lacks of a unique 
practical definition.  
The coupling {Hamiltonian} is usually described as the sum of an electrostatic interaction combined with a non-bonded term which is often modeled using Lennard-Jones force field to account for repulsion and dispersion effects\cite{SenThi-Ang-09,giovannini_general_2017,clemente_best_2023}.
{The Mixed cDFT-eDFT approach} advantageously replaces the exhaustive phase-space sampling required in QM/MM {molecular dynamics} with a self-consistent optimization 
of both classical and quantum functionals. While the idea of merging cDFT and quantum methods is completely natural its theoretical justification is not immediately straightforward.

Petrosyan and co-workers\cite{petrosyan_joint_2005,petrosyan_joint_2007} \manu{introduced the joint density functional theory 
which consists in a DFT formulation of the QM/MM problem.  
The authors obtain a multi component electron/nuclei/solvent DFT formulation,} which is
justified by the use of a Levy-Lieb constrained search procedure allowing to express the universal functional as depending 
on the electron density and the nuclear density rather than the full density matrix describing the electrons and nuclei. 
However, to derive a QM/MM formulation, they treat the environment particles as classical by '\textit{integrating out}' 
the electronic density associated with these nuclei, without explicitly indicating how this is done. 
Therefore, this step makes it difficult to pursue a rigorous derivation. Their working functional is instead defined as the sum of the 
Quantum Kohn-Sham functional describing the electrons of the solute, the classical functional describing the solvent and a coupling functional defined as the difference between the exact functional and the sum of the two former functionals. 
In other work\cite{JeaLevBor-JCTC-20, tang_solvent_2020}, a similar form of the functional was invoked without further theoretical justification.

Thus, while QM/cDFT approaches offer potentially attractive computational scaling \manu{and appears as a quite natural outcome}, 
their derivation from an exact theory remains somewhat ambiguous. 
This work addresses this gap by establishing a rigorous mathematical framework that provides a DFT-equivalent description of mixed QM/MM systems. 
Here, we focus on the canonical ensemble, deriving equations to compute the Helmholtz free energy 
using only the QM and MM one-body densities. 
Extending this formulation to the grand canonical ensemble for classical particles is conceptually straightforward but introduces 
additional complexity in the notations due to the varying number of particles. \manu{We therefore do not include it in the core of the paper, but rather in the supplementary materials.}  \gj{This derivation follows a similar route to its formally lighter canonical counterpart, which is reported in the present paper.} 
\gj{The main difference between the two ensembles is the existence of a unique mapping between the equilibrium density and the external potential in the grand canonical ensemble. 
In contrast, in the canonical ensemble, the equilibrium density is associated to a family of external potentials, 
defined up to an additive constant}\cite{gonzalez_density_1997,lutsko_classical_2022}. 

\manu{
Our derivation begins with two key assumptions:
(a) the entire system is partitioned into a QM region and a MM region; and 
(b) specific models are available for both the MM subsystem and the QM/MM coupling.
As such, we do not address the QM/MM crossover or the accuracy of the MM and QM/MM models themselves. 
We assume these models are given, as is standard in practical QM/MM calculations.

Under these assumptions, the primary objective of this paper is to derive, in a straightforward manner, 
the one-body density functional theory (DFT) formulation for a QM/MM system. Therefore, The key steps of this derivation are as follows:
i) We begin by \manulast{a qualitative discussion in Sec. \ref{sec:pre} sketching how the mathematical objects  
of QM/MM systems can be obtained by Wigner transformations,} and then define properly in Sec. \ref{sec:htot} 
the general form of the Hamiltonian of the system, considering a broadly applicable QM/MM Hamiltonian.
ii) We then express the exact solution of the problem using full-system quantities (\textit{i.e.}, those describing all particles). This includes 
 the Helmholtz free energy of the QM/MM system, along with the corresponding equilibrium density matrix and partition function 
(see Sec. \ref{sec:qmmm_ed}).
iii) Next, we reformulate the problem as the minimization of a functional over generic full-system density matrices. In our case, this means expressing the Helmholtz free energy as a functional of QM/MM density matrices that describe all particles in the system (see Sec. \ref{sec:exact_qmm_full}).
iv) We then decompose this functional into intrinsic (arising from kinetic energy, inter-particle interactions and entropic contributions) and external (arising from external potentials) contributions (see Sec  \ref{sec:rewrite_exact}).
v) Under certain assumptions about the form of the external potentials, we show how the external part of the functional can be written as a linear functional of one-body densities. In our case, this results in a functional of both classical and quantum one-body densities (see Sec. \ref{sec:levy-lieb}).
vi) Leveraging the many-to-one correspondence between full-system and one-body quantities, we perform a Levy–Lieb constrained search to rewrite the intrinsic free energy functional as a universal functional of the one-body densities (see Sec. \ref{sec:levy-lieb}).
vii) We then establish a connection between electronic DFT (eDFT) and classical DFT (cDFT) by introducing a universal QM/MM correlation functional of the one-body QM and MM densities (see Sec. \ref{sec:qmmm_expl}).
viii) Finally, we discuss this formulation in the context of solvation problems and recent developments in QM/cDFT coupling approaches (see Sec. \ref{sec:discuss}).
\manu{The Appendix \ref{sec:f_prop_proof} contains a proof of the variational properties of the QM/MM functional introduced here, and 
the Supplementary Material provides a recall of the Von-Neuman variational principle in statistical QM for the canonical ensemble, 
an alternative derivation of the QM/MM equilibrium density 
previously established in the adiabatic basis by Nielsen \textit{et. al.} in the context of QM/MM dynamics\cite{NieKapCic-JCP-01}, 
and an extension to the semi grand canonical ensemble of the present formalism. }
As this work is intended for both the QM and MM communities, which often differ in terminology and conceptual frameworks, we highlight certain elements throughout the text to aid accessibility. Some of these clarifications may appear self-evident to some readers; we hope, however, that they do not unduly burden the readability of the article.}

\section{Quantum-classical system in the Canonical ensemble: equilibrium density and variational formulation}
\label{eq:qmmm_1}
The objective of this section is to give a variational formulation for the Helmholtz free energy of a mixed $\qmb$/$\mmb$ 
system based on a functional minimization over general densities involving $\nqm$ and $\nmm$ variables. 
\manu{This is essentially a QM/MM equivalent to the Gibbs variational principle in classical statistical physics or the Von-Neumann variational principle 
in quantum statistical physics (see the supplementary information for a short review). 
This variational principle, although hardly usable in practice due to the high-dimensionality of the objects it manipulates, 
is an important milestone to properly obtain the one-body formulation, which is the aim of this paper.} 

\manulast{As the formalism uses mixed QM/MM objects that might seem rather odd in their mathematical design, 
we begin this derivation by providing in Sec.\ref{sec:pre} a qualitative discussion 
on how one can obtain these QM/MM objects thanks to the concept of Wigner transform. }
We then define in Sec. \ref{sec:htot} the working \gj{Hamiltonian that describes} the $\qmb$/$\mmb$ system. \gj{Again, we do not discuss the validity of the partitioning or the quality of the coupling term, we assume that these preliminary considerations have already been addressed. }
The equilibrium density of a mixed QM/MM system is then given in Sec. \ref{sec:qmmm_ed}, and we briefly discuss the physical meaning of that result. 
We eventually show in Sec. \ref{sec:exact_qmm} that the Helmholtz free energy can be obtained through a variational principle over mixed $\qmb$/$\mmb$ density matrices.

\subsection{\manulast{Qualitative discussion: dealing with both QM and MM particles thanks to the Wigner transform}}\label{sec:pre}
\manulast{
Properly defining objects for a QM/MM system can appear as \textit{a priori} relatively odd as the 
mathematical spaces of the two theories are very different. 
We provide here a qualitative discussion (which is fully expanded in the supplementary materials) on how 
one can obtain naturally QM/MM objects with the help of the so-called Wigner-Weyl transformation. 

Observables in classical statistical mechanics are obtained as integrals over the $(Q,P)$ classical phase space, 
\begin{equation}\label{eq:wigner_1}
 \langle O \rangle_f = \int dQdP f(Q,P) O(Q,P),
\end{equation}
where $f(Q,P)$ is a classical probability distribution and $O(Q,P)$ is a classical observable, 
while the QM theory uses density matrices $\hat{\rho}$ and potentially non local operators $\hat{O}$ 
acting in $\mathbb{R}\times \mathbb{R}$, namely  
\begin{equation}\label{eq:wigner_2}
 \langle \hat{H} \rangle_{\hat{\rho}} = \int dx dx' \hat{\rho}(x,x') \hat{O}(x',x), 
\end{equation}
where $\hat{\rho}(x,x')$ and $\hat{O}(x',x)$ are the real-space representation of the density matrix $\hat{\rho}$ 
and $\hat{O}$ operators, respectively. 
Nevertheless, one can notice that the $(Q,P)$ classical phase space is isomorph to the $(x,x')$ 
representations of the QM operators. The Wigner-Weyl transform\cite{Wigner-PR-32} exploits 
such a similarity and allows then for a phase space representation of QM density matrices and operators  
(see Ref. \onlinecite{Case-AJP-08} for a pedagogical introduction). 

More precisely, for any QM operator $\hat{O}$ one can define its Wigner transformation labelled $\tilde{O}_W$, 
which is now a \textit{function} of the $(Q,P)$ variables. 
One can also define the Wigner transformed density matrix $\tilde{\rho}_{W}(Q,P)$ 
associated to $\hat{\rho}$, which allows then to write the QM observable of Eq. \eqref{eq:wigner_2} 
as a classical phase space integral, namely
\begin{equation}
\langle \hat{O} \rangle_{\hat{\rho}} = \int dQdP \tilde{\rho}_{W}(Q,P) \tilde{O}_W(Q,P).
\end{equation}
An important property of this framework is that QM operators expressed as "simple" functions of 
the QM operators $\hat{Q}$ and $\hat{P}$ 
(\textit{i.e.} functions with no products of $\hat{P}$ and $\hat{Q}$) are transformed as 
the corresponding purely classical function. 
An important special case of this property is the Hamiltonian operator whose Wigner transformation is precisely 
the classical Hamiltonian, \textit{i.e.}  
\begin{equation}
 \tilde{H}_W(Q,P) = \frac{P^2}{2M} + V(Q)\equiv H(Q,P) .
\end{equation}
Therefore, the expectation value of the QM Hamiltonian over a QM density matrix is obtained as 
\begin{equation}
\langle \hat{H} \rangle_{\hat{\rho}} = \int dXdP \tilde{\rho}_{W}(Q,P) H(Q,P),
\end{equation}
which is very similar to the definition of the statistical classical mechanics. 
Nevertheless the quantum nature of the problem remains encoded in the fact that the object $\tilde{\rho}_{W}(Q,P)$ 
is potentially complex-valued in opposition to classical probability distributions.  

As shown in the seminal work of Wigner\cite{Wigner-PR-32}, its transformation applied to the QM equilibrium 
thermal density matrix (\textit{i.e.} $\hat{\rho}\propto \exp(-\beta \hat{H})$) 
is, at zeroth-order in $\hbar$, the usual classical Boltzmann distribution function 
(\textit{i.e.} $\tilde{\rho}_W\propto \exp(-\beta H(Q,P))$). 
The quantum nature of the problem appears then as higher order contributions expressed 
as powers of differential operators which vanish when 
$\hbar \rightarrow 0$, $\beta \rightarrow 0$, and/or $M\rightarrow \infty$, 
where $M$ is the mass of the QM particle. 
The Wigner transformation can therefore be used to naturally recover a classical description of a system 
in a semi classical limit governed by the parameters of the system such as the temperature or the mass of the particles. 

In the case now where two types of particles are present in the system with a clear difference of masses, 
one can perform a Wigner transformation only on the heavy particles and one obtains a power series 
in terms of the ratio of the masses $\alpha \ll 1$. 
As shown originally by Nielsen \textit{et. al.}\cite{NieKapCic-JCP-01}, the zeroth-order term in $\alpha$
yields to a classical description of the heavy particles, while maintaining the full QM description of the light 
particles. The obtained semi-Wigner transformed density matrix corresponds then to a QM/MM description of the system: 
the heavy particles are described by the classical variables $(Q,P)$ while the light particles 
are described by a non local operator. 
An important aspect of these results is that the QM operators depend parametrically on the $(Q,P)$ classical variables. 
We will use this important result as the starting of our derivation (see Sec. \ref{sec:qmmm_ed}) 
which will allow us to understand how generic QM/MM density matrices can be built in order to design our variational 
principle. 
}

\subsection{The Hamiltonian of the quantum-classical system}
\label{sec:htot}
%
\gj{We  consider a QM/MM system, where the partition and the choice of the model to describe the MM part and the coupling term have already been made} 

\gj{    We adopt the following notation conventions:
(i) lowercase letters denote purely quantum quantities (variables or operators), while uppercase letters denote classical or mixed quantum–classical quantities;
(ii) bold symbols denote collective quantities (over all particles), while non-bold symbols denote single-particle or pairwise quantities.
The only exception is the number of quantum particles, labeled  $\nqm$. Atomic units are used throughout.}


The total system is divided into a set of $\nmm$ identical classical particles, 
described by their classical positions $\boldqb=\{\varqb{1},\hdots,\varqb{\nmm}\}$ 
and linear momentum $\boldpb=\{\varpb{1},\hdots,\varpb{\nmm}\}$ 
and a set of $\nqm$ identical quantum particles described by their position variables 
$\boldqs=\{\varqs{1},\hdots,\varqs{\nqm}\}$.

\gj{The total Hamiltonian is written as the sum of the QM and MM Hamiltonians, and a QM/MM coupling term, namely}
\begin{equation}
 \label{eq:fulh_split}
 \fulh(\boldqb,\boldpb,\boldqs) = \hathqm(\boldqs) + \hmm(\boldqb,\boldpb) + \hatboldwqmmm(\boldqb,\boldqs).
\end{equation}
\manulast{Notice that the form of the QM/MM Hamiltonian as written in Eq. \eqref{eq:fulh_split} can actually be obtained as the partial Wigner 
transformation of a purely QM Hamiltonian, and where the Wigner transformation is done only on the $\nmm$ particles.} 
In the rest of the paper, explicit dependency on the $\qmb$ and $\mmb$ variables will be dropped for concision when unambiguous.
In Eq.\eqref{eq:fulh_split}, $\hathqm$ is the quantum Hamiltonian operator describing $\nqm$ quantum identical particles of mass $m$ (typically electrons). \gj{These particles 
interact with an external potential $\vextqm(\varqs{})$, which is local in the position representation, and with each other through a pairwise potential $\wqm(\varqs{i},\varqs{j})$. }
The explicit form of $\hathqm$ reads then 
\begin{equation}
\label{eq:hqm}
 \hathqm = \hatboldts + \hatboldvextqm + \hatboldwqm,
\end{equation}
with 
\begin{equation}
 \begin{aligned}
  &\hatboldts =  \sum_{i=1}^{\nqm} \frac{\hatps{i}^{2}}{2 m} ,\qquad 
   \hatps{i} = -i \nabla_{\varqs{i}}, 
 \end{aligned}
\end{equation}
\begin{equation}
 \label{eq:hqm_split}
 \hatboldvextqm = \sum_{i=1}^{\nqm} \vextqm(\varqs{i}),\qquad \hatboldwqm = \sum_{j=1}^{\nqm}\sum_{i>j} \wqm(\varqs{i},\varqs{j}).
 \end{equation}
Then, $\hmm$ is the classical \manu{model} Hamiltonian function representing $\nmm$ identical  particles of mass $M$, 
in interaction with a local external potential $\vextmm(\varqb{})$ and interacting among themselves through
a local pairwise potential $\wmm(\varqb{i},\varqb{j})$. 
\gj{These  are typically effective particles, such as atoms or rigid molecules, and they interact with each other through parametrized force fields. }
The explicit form of the classical Hamiltonian $\hmm$ reads then 
\begin{equation}
 \hmm = \boldtb + \boldvextmm + \boldwmm, 
\end{equation}
with 
\begin{equation}
 \label{eq:def_t_cl}
 \boldtb  = \sum_{i=1}^{\nmm} \frac{\varpb{i}^2}{2 M} ,  \qquad \boldvextmm = \sum_{i=1}^{\nmm} \vextmm(\varqb{i}),
\end{equation}
\begin{equation}
  \boldwmm= \sum_{j=1}^{\nmm}\sum_{i>j} \wmm(\varqb{i},\varqb{j}). 
\end{equation}
Regarding the coupling term $\wqmmm$, we simply assume here that it is an additive pairwise potential, \textit{i.e.}  
\begin{equation}
 \hatboldwqmmm  = \sum_{i=1}^{\nqm} \sum_{j=1}^{\nmm} \wqmmm(\varqs{i},\varqb{j})
\end{equation}
The function $\wqmmm(\qs,\qb)$ represents the effective interaction between 
an effective particle (such as a nuclei, an atom or a molecule) at position $\qb$ and a fundamental particle (such as an electron) 
at position $\qs$. 

\subsection{Mixed quantum-classical equilibrium density, partition function and Helmholtz free energy}
\label{sec:qmmm_ed}
\gj{We start by defining the exact equilibrium density matrix (EDM) of the QM/MM system in the canonical ensemble. This quantity is the analogue of the equilibrium density matrix of a purely quantum system or, equivalently, of the equilibrium probability density of a purely classical (MM) system.}

\gj{With the  Hamiltonian of Eq. \eqref{eq:fulh_split}, the EDM takes the following form}

\begin{equation}
 \label{eq:eq_qmmm_rho_p}
 \begin{aligned}
 \mathcal{M}(\boldqb,\boldpb,\boldqs,\boldqps) & = \frac{\elemm{\boldqs}{e^{-\beta \fulh(\boldqb,\boldpb) }}{\boldqps}}{\eqztot}\\
 & =  \frac{\elemm{\boldqs}{e^{-\beta(\hathqm + \hatboldwqmmm(\boldqb))}}{\boldqps}e^{-\beta\hmm(\boldqb,\boldpb)}}{\eqztot}.
 \end{aligned}
 \end{equation}
 In Eq.~\eqref{eq:eq_qmmm_rho_p} $\beta=T^{-1}$ \manu{in atomic units}, \manu{and $\ket{\boldqs}$ and $\ket{\boldqps}$ are  position eigenstates for the $\nqm$ quantum variables}. 
The dependency of $\hatboldwqmmm$ on the classical variables, $\boldqb$, has been made explicit.
\gj{Because  Eq. \eqref{eq:eq_qmmm_rho_p} involves the exponential of a quantum operator, 
the QM/MM EDM is itself a quantum operator, namely a quantum density matrix (\textit{i.e.} positive definite and of trace  unity). 
However, it is parametrized by the classical variables $(\boldqb,\boldpb)$  and therefore
the QM system depends on the configuration of the classical particles. 
To understand the physical meaning of the EDM, 
it is instructive to analyse its diagonal elements, \textit{i.e.}} when $\boldqs=\boldqps$. 
The quantity $\mathcal{M}(\boldqb,\boldpb,\boldqs,\boldqs)$ gives the probability of finding the quantum particles at positions $\boldqs =\{\varqs{1},\hdots,\varqs{\nqm}\}$ 
while the classical particles have positions $\boldqb=\{\varqb{1},\hdots,\varqb{\nmm}\}$ and momenta $\boldpb=\{\varpb{1},\hdots,\varpb{\nmm}\}$. 
\gj{Off-diagonal elements encode purely quantum effects, such as nonlocality, and are necessary to compute the action of quantum operators—for example, the kinetic energy operator, which involves a Laplacian in the quantum subsystem.}

A \manu{pictorial} illustration of the physical meaning of the equilibrium density of Eq. \eqref{eq:eq_qmmm_rho_p} 
is provided by the Born-Oppenheimer description of a molecule, \textit{i.e.} QM electrons and classical nuclei. 
In that case, $\hathqm$ is composed of the kinetic operator of the electrons 
and their mutual coulomb interaction, with no external potential  for either the $\qmb$ or $\mmb$ parts. 
The operator $ \hatboldwqmmm(\boldqb)$ represents the nuclei-electron attraction, 
while $\boldwmm(\boldqb)$ represents the Coulomb repulsion between the nuclei. 
Thus, $\fulh(\boldqb,\boldpb)$ entering in the exponential of Eq. \eqref{eq:eq_qmmm_rho_p} corresponds the Born-Oppenheimer Hamiltonian. 

From a more formal point of view, it is worth noticing that   Eq. \eqref{eq:eq_qmmm_rho_p} corresponds to the real-space representation 
of the EDM, originally obtained by Nielsen \textit{et. al.}\cite{NieKapCic-JCP-01} in the adiabatic basis,  
\manu{\textit{i.e.} in the basis of eigenvectors of $\hathqm + \hatboldwqmmm(\boldqb)$. 
In this basis, the density matrix is diagonal , as found by Nielsen \textit{et. al.}\cite{NieKapCic-JCP-01}. }
\manu{In the case of quantum electrons and classical nuclei, the adiabatic states are  the eigenstates of the Born-Oppenheimer Hamiltonian.} 
Therefore, Eq. \eqref{eq:eq_qmmm_rho_p} ignores the non-adiabatic couplings between these  states. 
In the context of \textit{ab initio} MD, this implies that $\mathcal{M}(\boldqb,\boldpb,\boldqs,\boldqps)$ can be obtained from  classical nuclear trajectories, with forces averaged over all adiabatic electronic states weighted by the corresponding Boltzmann factors.
Consequently, it entirely ignores surface hopping between different adiabatic states, which is why $\mathcal{M}(\boldqb,\boldpb,\boldqs,\boldqps)$ 
is referred to as the \textit{adiabatic} QM/MM density matrix. 
There is a flourishing literature addressing methods going beyond this adiabatic approximation, specially in the context of quantum dynamics or statistical mechanics for QM/MM systems (see for instance Ref. \onlinecite{Kapral-JPCM-15} for a review), but  treatment of non adiabatic couplings is beyond the scope of this paper. 

Although the form of Eq. \eqref{eq:eq_qmmm_rho_p} follows the usual $\exp(-\beta \hat{H})$ form, 
it can be obtained from first principle by starting from a Hamiltonian where both sets of particles 
are considered to be quantum, \manu{\textit{i.e.} with their kinetic energy propotional to the Laplacian operator}. 
\manulast{The main idea, sketched in Sec. \ref{sec:pre},} is to then perform a Wigner transform over the $\nmm$ particles only and eventually a semi-classical limit is obtained by letting the ratio of masses $(m/M)^{1/2}$ tends towards 0. Conserving only the zeroth order term of this semi-classical limit yields to Eq. \eqref{eq:eq_qmmm_rho_p}. 
This expansion in terms of the ratio of the masses is natural since the quantum particles, typically electrons, are usually much lighter than the classical ones, typically atoms or molecules. 
The original derivation of Nielsen \textit{et. al.}\cite{NieKapCic-JCP-01} used the Liouvillian imaginary time evolution equation of the density matrix together with the Poisson-brakets operators\cite{Wigner_bracket-67}, which is a well-known formalism in the quantum dynamics community\cite{KapCic-JCP-99,NieKapCic-JCP-01,Kapral-ARPC-06,Kapral-JPCM-15}.
However, we provide in the supplementary materials an alternative derivation which involves a similarity transformation and resorts to the Baker-Campbell-Hausdorf expansion to express the Wigner transform Hamiltonian as a series of nested commutators, which might appear more familiar to the "static" quantum chemistry community.

In Eq. \eqref{eq:eq_qmmm_rho_p}, the normalization corresponds to the partition function of the QM/MM system which is expressed as
\begin{equation}
 \label{eq:eq_qmmm_z_p}
 \begin{aligned}
  \eqztot  =  \frac{1}{\nmm! }&\intdqb \dpb \, e^{-\beta\hmm(\boldqb,\boldpb)} \\
    & \int\!\text{d}\boldqs 
    \bra{\boldqs}{e^{-\beta(\hathqm + \hatboldwqmmm(\boldqb))}}\ket{\boldqs}.
 \end{aligned}
\end{equation}
The corresponding Helmholtz free energy is therefore given by 
\begin{equation}
 \potcanotot = -\kb T \log(\eqztot).
\end{equation}
\manu{Any equilibrium property $ \langle O \rangle_{\text{eq}}$ can be obtained from derivatives of $\eqztot$.  
Alternatively, $ \langle O \rangle_{\text{eq}}$ 
can be computed as the trace of the product of $\mathcal{M}$ and the operator $\hat{O}$, which may depend on both quantum and classical variables} 
\manu{
\begin{equation}
 \begin{aligned}
 \langle O \rangle_{\text{eq}} 
  & = \intdqb \dpb \,\int\!\text{d}\boldqs \text{d}\boldqps \hat{O}(\boldqb,\boldpb,\boldqs,\boldqps) \mathcal{M}(\boldqb,\boldpb,\boldqps,\boldqs) \\
  & \equiv \text{Tr} \{ \hat{O} \hat{\mathcal{M}} \}.\\
 \end{aligned}
\end{equation}
}

Although the expression of the equilibrium density in Eq. \eqref{eq:eq_qmmm_rho_p} is relatively simple, \gj{it benefits from further} comments. 
Firstly, as usual in classical statistical mechanics, the contribution to the Helmholtz free energy coming from the purely classical kinetic term 
in $\hmm(\boldqb,\boldpb)$ can be computed analytically. 
Therefore, it is convenient to separate the equilibrium density according to its dependence on $\boldqb$ and $\boldpb$ 
\begin{equation}
 \label{eq:eq_qmmm_rho_p_2}
 \begin{aligned}
 \mathcal{M}(\boldqb,\boldpb,\boldqs,\boldqps) = \frac{e^{-\beta \boldtb}}{z_{\boldpb}} \eqrdm(\boldqb,\boldqs,\boldqps),
 \end{aligned}
\end{equation}
where  $\boldtb$ is the purely classical kinetic energy defined in Eq. \eqref{eq:def_t_cl}, and  
\begin{equation}
 \label{eq:eq_qmmm_rho}
 \begin{aligned}
\eqrdm(\boldqb,\boldqs,\boldqps) = 
\frac{\elemm{\boldqs}{e^{-\beta(\hathqm + \hatboldwqmmm(\boldqb))}}{\boldqps}e^{-\beta\vmm(\boldqb)}}{\eqz},
 \end{aligned}
\end{equation}
\begin{equation}
 \label{eq:vtotmm}
 \begin{aligned}
  \vmm & = \boldvextmm + \boldwmm, 
 \end{aligned}
\end{equation}
\begin{equation}
 \label{eq:eq_qmmm_z_p2}
 \begin{aligned}
  z_{\boldpb} & =  \frac{1}{\nmm!}\int\dpb e^{-\beta \boldtb}= \frac{1}{\nmm!}\bigg(\frac{2\pi M}{\beta } \bigg)^{\frac{3}{2}\nmm },
 \end{aligned}
\end{equation}
\begin{equation}
 \label{eq:eq_qmmm_z}
 \begin{aligned}
  \eqz & = \intdqb \intdqs  
    \elemm{\boldqs}{e^{-\beta(\hathqm + \hatboldwqmmm(\boldqb))}}{\boldqs}e^{-\beta\vmm(\boldqb)},
 \end{aligned}
\end{equation}
\begin{equation}
 \label{eq:eq_qmmm_z_tot}
 \begin{aligned}
  \eqztot & = z_{\boldpb}\eqz  ,
 \end{aligned}
\end{equation}
such that the free energy can be written as 
\begin{equation}
 \begin{aligned}
 \potcanotot&  = \potcanoz+F_{\boldpb} ,
 \end{aligned}
\end{equation}
where 
 \begin{equation}
 \label{eq:Fid}
 F_{\boldpb} = -\kb T\log(z_{\boldpb}),
 \end{equation} 
and 
\begin{equation}
 \label{eq:def_pot_cano_int}
\potcanoz = -\kb T\log(\eqz).
\end{equation}
\gj{$\potcanoz$ represents the contribution to the Helmholtz free energy of the QM/MM system arising from particle interactions.} 
\gj{Computing $\potcanoz$ from Eq.~\eqref{eq:def_pot_cano_int} is infeasible for realistic systems, as it requires evaluating Eq.~\eqref{eq:eq_qmmm_z}. The aim of this paper is to develop an alternative, more practical, route to $\potcanoz$.}

The EDM $\eqrdm(\boldqb,\boldqs,\boldqps)$ of Eq.\eqref{eq:eq_qmmm_rho} can be decomposed into the product of a purely quantum density matrix { $\eqrdmqm(\boldqs,\boldqps,\boldqb)$ (more precisely its matrix element evaluated in real space) and a purely 
classical function $ \eqrdmmm(\boldqb)$}, \textit{i.e.}
\begin{equation}
 \begin{aligned}
 \label{eq:Peq=f*rho}
 \eqrdm(\boldqb,\boldqs,\boldqps) = \eqrdmqm(\boldqs,\boldqps,\boldqb) \eqrdmmm(\boldqb),
 \end{aligned}
\end{equation}
with 
\begin{equation}
 \label{eq:eq_rdmqm}
 \begin{aligned}
 \eqrdmqm(\boldqs,\boldqps,\boldqb) & =\elemm{\boldqs}{\hateqrdmqm(\boldqb)}{\boldqps},
 \end{aligned}
\end{equation}
\begin{equation}
 \label{eq:eq_rdmqm_2}
 \begin{aligned}
  \hateqrdmqm(\boldqb) = \frac{1}{\eqzqm(\boldqb)}e^{-\beta(\hathqm + \hatboldwqmmm(\boldqb))},
 \end{aligned}
\end{equation}
\begin{equation}
 \label{eq:eq_qm_z}
 \begin{aligned}
  \eqzqm(\boldqb) & = \intdqs  
    \elemm{\boldqs}{e^{-\beta(\hathqm + \hatboldwqmmm(\boldqb))}}{\boldqs},
 \end{aligned}
\end{equation}
\begin{equation}
 \label{eq:eq_rdmmm}
 \begin{aligned}
 \eqrdmmm(\boldqb) = \frac{\eqzqm(\boldqb)}{\eqz}e^{-\beta\vmm(\boldqb)}. 
 \end{aligned}
\end{equation}
$\eqz$ and $\eqzqm(\boldqb)$ are related through the following equation
\begin{equation}
 \label{eq:eq_qmmm_z_bis}
 \begin{aligned}
  \eqz & = \intdqb \eqzqm(\boldqb)e^{-\beta\vmm(\boldqb)}.
 \end{aligned}
\end{equation}
It is noteworthy that, \manu{for each value of $\boldqb$}, $\hateqrdmqm(\boldqb)$ defined in Eq. \eqref{eq:eq_rdmqm} and \eqref{eq:eq_rdmqm_2} is a 
density matrix, \manu{\textit{i.e.} positive definite and of unity norm.} 
Similarly, $\eqrdmmm$ of Eq. \eqref{eq:eq_rdmmm} is a probability distribution. 
\gj{Therefore, $\hateqrdmqm$ and $\eqrdmmm$ are  special cases of more general 
objects: density matrices representing $\nqm$ quantum particles and probability distribution functions representing $\nmm$ classical particles, respectively.} 
This will allow  to formulate a variational formulation of the Helmholtz free energy for a QM/MM system in Sec. \ref{sec:exact_qmm}.

{Last but not least, the decomposition of the equilibrium density of Eq. \eqref{eq:Peq=f*rho} into a product of a purely quantum equilibrium density and a
purely classical equilibrium density} might suggest  that the quantum and classical subsystems are uncorrelated, even though they interact.
Nevertheless, 
the quantum part of the equilibrium density, $\hateqrdmqm$,  depends on the coordinates of the classical 
systems through the $\qmb$/$\mmb$ interaction potential $\hatboldwqmmm(\boldqb)$. \gj{The classical part, $\eqrdmmm$, is in turn affected by the quantum degrees of freedom through its normalization, $\eqzqm(\boldqb)$. To see this more explicitly, one can notice that the form of Eq. \eqref{eq:eq_rdmmm} can be rewritten as} 
\begin{equation}
 \label{eq:eq_rdmmm_bis}
 \begin{aligned}
 \gj{\eqrdmmm(\boldqb) = \frac{1}{\eqz}e^{-\beta\big(\vmm(\boldqb) + U_{\text{eff}}(\boldqb)\big)},}
 \end{aligned}
\end{equation}
\gj{where now the classical particles experience an additional the effective potential $U_{\text{eff}}(\boldqb)$ due to the 
presence of the QM system, which is defined as}
\begin{equation}
 \gj{U_{\text{eff}}(\boldqb) = -\frac{1}{\beta}\log\big(\eqzqm(\boldqb)\big).}
\end{equation}
Therefore, the two systems are  correlated.

\subsection{Variational formulation of the Helmholtz free energy for a $\qmb$/$\mmb$ system}
\label{sec:exact_qmm}
\gj{Both classical and quantum DFT can be formulated through  the Levy-Lieb constrained-search formalism. 
The first step is to express  the quantum ground state energy and the classical Helmholtz free energy 
as a general functional minimization problem, defined over the many-body functions, namely the electronic density matrix and the probability distribution, respectively. 

Since the goal of this work is to obtain a DFT formulation of the QM/MM problem, the first step is to reformulate the Helmholtz free energy 
of Eq. \eqref{eq:def_pot_cano_int} as a functional minimization problem over QM/MM density matrices whose form generalize the EDM of Eq.~\eqref{eq:Peq=f*rho}.} This is the aim of this section. 
To the best of our knowledge, this was never formalized before.

\subsubsection{Variational formulation}
\label{sec:exact_qmm_full}
As we know that the equilibrium density for a mixed $\qmb$/$\mmb$ system is expressed as a special case of product of a quantum density matrix by a classical probability distribution, 
we can now reformulate the Helmholtz free energy $\potcanoz$ as a variational problem over QM and MM objects, 
such that the equilibrium density of Eq. \eqref{eq:Peq=f*rho} is the minimizer of that functional and that it yields 
the correct Helmholtz free energy of Eq. \eqref{eq:def_pot_cano_int}. 

Thus, we would like to write the Helmholtz free energy as the minimum of a functional defined on a couple of quantum and classical objects as follows 
\begin{equation}
 \label{eq:def_e_cano_eq}
 \potcanoz = \min_{\hatrdmqm,\rdmmm} \ecano[\hatrdmqm,\rdmmm] ,
\end{equation}
where the operator $\hatrdmqm$ is a quantum density matrix representing $\nqm$ particles 
but \textit{also parametrically depending on all classical variables $\boldqb=\{\varqb{1},\hdots,\varqb{\nmm}\}$}, \textit{i.e.}
\begin{equation}
 \hatrdmqm \equiv \hatrdmqm(\boldqb),
\end{equation}
and $\rdmmm$ is a classical probability density representing $\nmm$ classical particles and not depending 
on any quantum variable. 
{Therefore, $\hatrdmqm$ and $\rdmmm$ must be normalized to unity in order to be acceptable density matrices and probability measures, respectively, \textit{i.e} } 
\begin{equation}
 \label{eq:norm_cond}
 \begin{aligned}
 &\intdqs\,\,\rdmqm(\boldqs,\boldqs,\boldqb) = 1\quad \forall\,\, \boldqb,
 &\intdqb\,\,\rdmmm(\boldqb) = 1,
 \end{aligned}
\end{equation}
where $\rdmqm(\boldqs,\boldqps,\boldqb)$ is the real-space representation of the operator $\hatrdmqm$,
\textit{i.e.}
\begin{equation}
 \begin{aligned}
 \rdmqm(\boldqs,\boldqps,\boldqb) = \elemm{\boldqs}{\hatrdmqm(\boldqb)}{\boldqps},
 \end{aligned}
\end{equation}
such that that the product of the two functions is normalized to unity
\begin{equation}
 \label{eq:norm_prod_rhop}
  \intdqb\intdqs\,\,\rdmqm(\boldqs,\boldqs,\boldqb)\rdmmm(\boldqb)=1.
\end{equation}
{Therefore, the equilibrium density of the mixed QM/MM system described in Eq. \eqref{eq:Peq=f*rho} lies within the functional space {defined by all the couple of functions and density matrices satisfying Eqs. \eqref{eq:norm_cond}-\eqref{eq:norm_prod_rhop}}.}
As shown in the appendix (see section \ref{sec:f_prop_proof}), if we define the Helmholtz functional as follows
\begin{equation}
 \label{eq:pot_cano}
 \begin{aligned}
  \ecano[\hatrdmqm,\rdmmm] = \uint[\hatrdmqm,\rdmmm] - \kb T \entrop[\hatrdmqm,\rdmmm],
 \end{aligned}
\end{equation}
where $\uint$ is an energy functional excluding the classical kinetic term (which has already been accounted for in $F_{\boldpb}$) 
\begin{equation}
 \uint[\hatrdmqm,\rdmmm] = \tr\big\{\hatrdmqm \rdmmm(\hathqm+ \hatboldvextqm + \boldwmm+\boldvextmm + \hatboldwqmmm)\big\},
\end{equation}
and $\entrop$ is the dimensionless entropy of the total system 
\begin{equation}
 \entrop[\hatrdmqm,\rdmmm] = -\tr\big\{\hatrdmqm\rdmmm\log\big(\hatrdmqm\rdmmm \big)\big\},
\end{equation}
then the only minimizer of the functional of Eq. \eqref{eq:pot_cano} is  $(\hateqrdmqm,\eqrdmmm)$  
 of Eq. \eqref{eq:Peq=f*rho}, \textit{i.e}
\begin{equation}
 \begin{aligned}
 - \kb T\log(\eqz) \le \ecano[\hatrdmqm,\rdmmm]\quad \forall \,\,(\hatrdmqm,\rdmmm), 
 \end{aligned}
\end{equation}
and 
\begin{equation}
 \begin{aligned}
 \ecano[\hatrdmqm,\rdmmm] = F_0 = - \kb T\log(\eqz) \Leftrightarrow (\hatrdmqm,\rdmmm)=(\hateqrdmqm,\eqrdmmm).
 \end{aligned}
\end{equation}
\manu{This variational formulation, is the analogue of the Gibbs or Von Neumann variational principles, but applied to a QM/MM system.} 
\gj{As with any variational formulation, it naturally enables the construction of approximations by imposing constraints on the variables of minimization, the archetypal example being the mean-field approach, which is briefly discussed in Sec. \ref{sec:discuss}.}

\subsubsection{Rewriting of the functional $\ecano[\hatrdmqm,\rdmmm]$}
\label{sec:rewrite_exact}
Because the function $\rdmmm$ does not \manu{explicitly} depend  on the quantum variables and because we chose the normalization conditions 
of Eq. \eqref{eq:norm_cond}, we can naturally decompose the entropy as 
\begin{equation}
 \begin{aligned}
  \entrop[\hatrdmqm,\rdmmm] & = S_{\qm}^{\mm}[\hatrdmqm,\rdmmm] + S_{\mm}[\rdmmm],
 \end{aligned}
\end{equation}
where $S_{\mm}[\rdmmm]$ is the purely classical dimension-less entropy 
\begin{equation}
 \label{eq:s_mm}
 \begin{aligned}
  S_{\mm}[\rdmmm]& = -\intdqb\,\rdmmm(\boldqb) \log(\rdmmm(\boldqb)),
\end{aligned}
\end{equation}
and $S_{\qm}^{\mm}[\hatrdmqm,\rdmmm] $ is the quantum-classical dimension-less entropy 
\begin{equation}
 \label{eq:s_qmmm}
 \begin{aligned}
 S_{\qm}^{\mm}[\hatrdmqm,\rdmmm] & = -\tr\big\{\hatrdmqm\rdmmm\log(\hatrdmqm)\big\}\\ 
               & = -\intdqb\,\rdmmm(\boldqb)\intdqs\,\rdmqm(\boldqs,\boldqs,\boldqb) \log(\rdmqm(\boldqs,\boldqs,\boldqb)), \\
 \end{aligned}
\end{equation}
which depends explicitly on both $\hatrdmqm$ and $\rdmmm$. 

We can similarly decompose $\uint$ as follows 
\begin{equation}
 \begin{aligned}
  \uint[\hatrdmqm,\rdmmm] &=  E_{\mm}[\rdmmm] + E_\mm^{\text{ext}}[\rdmmm] \\
    &+ E_\qm[\hatrdmqm,\rdmmm] +E_\qm^{\text{ext}}[\hatrdmqm,\rdmmm] 
                          + E_{\qm}^{\mm}[\hatrdmqm,\rdmmm],
 \end{aligned}
\end{equation}
where $E_\mm[\rdmmm]$ is the mutual interaction of the classical part
\begin{equation}
 \label{eq:def_emm}
 \begin{aligned}
  E_\mm[\rdmmm] &=\intdqb \rdmmm(\boldqb) \boldwmm(\boldqb), \\
 \end{aligned}
\end{equation}
$E_\mm^{\text{ext}}[\rdmmm]$ is {the functional accounting for the interaction of the classical particles with the external potential}
\begin{equation}
 \label{eq:def_vmmext}
 \begin{aligned}
  E_\mm^{\text{ext}}[\rdmmm]&=\intdqb \rdmmm(\boldqb) \boldvextmm(\boldqb), \\
 \end{aligned}
\end{equation}
$E_\qm[\hatrdmqm,\rdmmm]$ is the kinetic and mutual interaction of the quantum part 
\begin{equation}
 \begin{aligned}
  E_\qm&[\hatrdmqm,\rdmmm]=\tr\{(\hatboldts + \hatboldwqm) \rdmmm\hatrdmqm\} \\
                  &=\intdqb \rdmmm(\boldqb) \\ 
 & \quad \intdqs \dqps  (\bf{t}(\boldqs,\boldqps) + \delta(\boldqs-\boldqps)\wqm(\boldqs,\boldqps))\rdmqm(\boldqs,\boldqps,\boldqb) ,
 \end{aligned}
\end{equation}
$E_\qm^{\text{ext}}[\hatrdmqm,\rdmmm]$ is {the functional accounting for the interaction of the quantum particles with the external potential} 
\begin{equation}
 \label{eq:def_vqmext}
 \begin{aligned}
  E_\qm^{\text{ext}}[\hatrdmqm,\rdmmm]&=\tr\{\hatboldvextqm \rdmmm\hatrdmqm\} \\
                  &=\intdqb \rdmmm(\boldqb) \intdqs  \boldvextqm (\boldqs,\boldqs) \rdmqm(\boldqs,\boldqs,\boldqb),
 \end{aligned}
\end{equation}
and eventually $ E_{\qm}^{\mm}[\hatrdmqm,\rdmmm]$  is the quantum-classical interaction,
\begin{equation}
 \label{eq:e_qmmm}
 \begin{aligned}
 E_{\qm}^{\mm}&[\hatrdmqm,\rdmmm] = \tr\{\hatboldwqmmm \rdmmm \hatrdmqm\}\\
               &= \intdqb\rdmmm(\boldqb) \intdqs \intdqps \boldwqmmm(\boldqb,\boldqs,\boldqps)\rdmqm(\boldqps,\boldqs,\boldqb).
 \end{aligned}
\end{equation}
It is   \gj{worth noting that the form of Eq. \eqref{eq:e_qmmm} is quite general, since $\boldwqmmm(\boldqb,\boldqs,\boldqps)$ is \textit{a priori} non local.} This form of interaction includes, for example, Lennard-Jones potentials, frozen density  Coulomb interaction 
or density overlap models for Pauli repulsion\cite{WhePri-MP-90}.


Eventually, once all terms are inserted in Eq. \eqref{eq:def_e_cano_eq}, the Helmholtz free energy $\potcanoz$ can be obtained as 
\begin{equation}
 \begin{aligned}
 \label{eq:def_e_cano_eq_2}
 \potcanoz = \min_{\hatrdmqm,\rdmmm} \big\{ &E_\qm[\hatrdmqm,\rdmmm]   +E_\qm^{\text{ext}}[\hatrdmqm,\rdmmm] + E_{\mm}[\rdmmm] + E_\mm^{\text{ext}}[\rdmmm] \\ 
                 &+ E_{\qm}^{\mm}[\hatrdmqm,\rdmmm] -\kb T \big(S_{\qm}^{\mm}[\hatrdmqm,\rdmmm] + S_{\mm}[\rdmmm]\big)\big\}.
 \end{aligned}
\end{equation}
\gj{Eq. \eqref{eq:def_e_cano_eq_2} therefore provides    a variational formulation of $\potcanoz$ for a mixed $\qmb$/$\mmb$ system, 
expressed as a minimization over the couple $(\hatrdmqm,\rdmmm)$.} 
One can see from Eqs. \eqref{eq:s_mm}, \eqref{eq:def_emm} and \eqref{eq:def_vmmext} 
that the functionals corresponding to the $\mmb$ part of the energy and entropy \manu{(\textit{i.e.} $E_\mm^{\text{ext}}[\rdmmm]$, $E_{\mm}[\rdmmm]$ and $S_{\mm}[\rdmmm]$)}  
are, \textit{per se}, completely independent of the quantum part of the system. 
\gj{They  only depend on the classical probability $\rdmmm$.
Nevertheless, the quantum and classical parts remain coupled in the minimization  of Eq. \eqref{eq:def_e_cano_eq_2} through the coupling functional $E_{\qm}^{\mm}[\hatrdmqm,\rdmmm]$.
In contrast, the  $\qmb$ parts  energy and entropy functionals
explicitly depend on the classical $\mmb$ probability distribution $\rdmmm$.} 

{Although Eq. \eqref{eq:def_e_cano_eq_2} provides a variational formulation 
of the Helmholtz free energy for a QM/MM system,} the minimization variables $(\hatrdmqm,\rdmmm)$ remain very high-dimensional \manu{because they 
 explicitly describe every individual particle in the system}.
\gj{The objective of the following section is to reformulate the variational principle of  Eq. \eqref{eq:def_e_cano_eq_2} 
using functionals depending on much lower-dimensional objects, \textit{i.e.} the one-body QM and MM densities}.

\section{One-body density formulation through Levy-Lieb constrained search formulation}
\label{sec:edft-cft}
\manu{Starting from the variational formulation of the Helmholtz free energy for the QM/MM system in terms of the 
$\nqm+\nmm$ quantities in  Eq. \eqref{eq:def_e_cano_eq_2}, we want to obtain a one-body DFT formulation. As in classical and quantum DFT, the key idea is 
to split the total energy into an \textit{external part} and an \textit{intrinsic part}.} 
More precisely, \gj{because of the one-body nature of both classical and quantum \textit{external} potentials $\vextmm(\qb)$ and $\vextqm(\qs)$, and the permutational symmetry of both $\rdmmm$ and $\hatrdmqm$, the  corresponding energy terms 
$E_\mm^{\text{ext}}[\rdmmm]$ and $E_\qm^{\text{ext}}[\hatrdmqm,\rdmmm]$ in Eqs. \eqref{eq:def_vmmext} and \eqref{eq:def_vqmext}  
can be expressed in terms of the one-body densities. } 
Then, by applying the Levy-Lieb constrained search, we can also reformulate the \manu{\textit{intrinsic} part of the energy as a functional 
of the one-body quantum and classical densities.}  
Therefore, the variational problem of Eq. \eqref{eq:def_e_cano_eq_2} is directly expressed as a functional of the one-body QM and MM densities rather than in terms of full $\nmm+\nqm$ variables quantities.
\subsection{Classical and quantum one-body density and the Levy-Lieb constrained search}
\label{sec:levy-lieb}
The classical one-body density is naturally defined as the trace over all quantum variables and all 
classical variables but one of the product $\hatrdmqm(\boldqs,\boldqb)\rdmmm(\boldqb)$, 
which, thanks to the normalization of Eq. \eqref{eq:norm_cond}, is simply 
\begin{equation}
 \begin{aligned}
 \label{eq:def_rdm_mm_2}
 \densmm(\qb)  = \nmm\int\text{d}\varqb{2}\mydots \text{d}\varqb{\nmm} \rdmmm(\qb,\varqb{2},\mydots,\varqb{\nmm}). 
 \end{aligned}
\end{equation}
Similarly we can define the QM one-body density $\densqm(\qs)$ as 
\begin{equation}
 \begin{aligned}
 \label{eq:def_rdm_qm_1}
  \densqm(\qs)& = \nqm \int \dqss_2 \mydots \dqss_{\nqm} \int \dqb \rdmqm(\qs,\qs_2,\mydots,\qs,\qs_2,\boldqb) \rdmmm(\boldqb).\\
 \end{aligned}
\end{equation}

\gj{Thanks to the permutational symmetry of both $\hatrdmqm(\boldqb)$ and $\rdmmm(\boldqb)$ with respect to the 
exchange of either QM or MM coordinates,  the two external functionals can be rewritten  as
functionals of the QM and MM one-body densities, respectively}
\begin{equation}
 \label{eq:def_vmmext_2}
 \begin{aligned}
  E_\mm^{\text{ext}}[\rdmmm] &=\intdqb \rdmmm(\boldqb) \boldvextmm(\boldqb) \\
                             &=\int\text{d}\qb \densmm(\qb) \vextmm(\qb) \\
                             &\equiv (\vextmm|\densmm),
 \end{aligned}
\end{equation}
\begin{equation}
 \label{eq:def_vqmext_2}
 \begin{aligned}
 E_\qm^{\text{ext}}[\hatrdmqm,\rdmmm] &=\intdqb \rdmmm(\boldqb) \intdqs  \boldvextqm (\boldqs,\boldqs) \rdmqm(\boldqs,\boldqs,\boldqb) \\
 & = \intdqss \densqm(\qs)\vextqm(\qs)\\
                  & \equiv (\vextqm|\densqm).
 \end{aligned}
\end{equation}

Since the contributions arising from the external potentials can be  directly expressed  in terms of the
one-body densities $(\densqm,\densmm)$, we now  introduce a universal functional of $(\densqm,\densmm)$ 
which contains all the information required to compute the  free energy. 
This functional is built using the Levy-Lieb constrained search formalism, as in eDFT. 
First, we introduce the \textit{intrinsic} Helmholtz free energy functional of the $N$-body densities
\begin{equation}
 \label{def:levy_loc}
 \begin{aligned}
 f[\hatrdmqm,\rdmmm] 
 & = E_{\mm}[\rdmmm] + E_\qm[\hatrdmqm,\rdmmm] + E_{\qm}^{\mm}[\hatrdmqm,\rdmmm] - \kb T \entrop[\hatrdmqm,\rdmmm].
 \end{aligned}
\end{equation}
such that we can then rewrite the Helmholtz free energy as 
\begin{equation}
 \label{eq:def_e_cano_eq_ll}
 \potcanoz = \min_{\hatrdmqm,\rdmmm} 
 \big( E_\qm^{\text{ext}}[\hatrdmqm,\rdmmm] + E_\mm^{\text{ext}}[\rdmmm] + 
f[\hatrdmqm,\rdmmm] \big).
\end{equation}
Then, the many-to-one correspondence between $(\hatrdmqm,\rdmmm)$ and $(\densqm,\densmm)$ allows us to introduce 
the Levy-Lieb constrained search as follows 
\begin{equation}
 \label{eq:ll_0}
 \begin{aligned}
   \min_{\hatrdmqm,\rdmmm} \ecano[\hatrdmqm,\rdmmm] = \min_{(\densqm,\densmm)} \,\,
 \min_{(\hatrdmqm,\rdmmm)\rightarrow (\densqm,\densmm)} \ecano[\hatrdmqm,\rdmmm],
 \end{aligned}
\end{equation}
where the notation $(\hatrdmqm,\rdmmm)\rightarrow (\densqm,\densmm)$ means that the search is performed over 
the set of couple of $\nqm+\nmm$-body densities $(\hatrdmqm,\rdmmm)$ that yield the 1-body densities  
$(\densqm,\densmm)$.  
Therefore, by remembering that the external parts depends only on the one-body densities 
(see Eqs. \eqref{eq:def_vmmext_2} and \eqref{eq:def_vqmext_2}), 
one can rewrite Eq. \eqref{eq:ll_0} as
%
\begin{equation}
 \label{eq:ll_2}
 \begin{aligned}
   \min_{\hatrdmqm,\rdmmm} \ecano[\hatrdmqm,\rdmmm] = 
   &\min_{(\densqm,\densmm)} \big[ (\vextmm|\densmm) + (\vextqm|\densqm) \\
  +&  
 \min_{(\hatrdmqm,\rdmmm)\rightarrow (\densqm,\densmm)} f[\hatrdmqm,\rdmmm] \big].
 \end{aligned}
\end{equation}
\gj{We have just introduced} the universal Levy-Lieb functional of the one-body classical and quantum density,
\begin{equation}
 \label{eq:def_ll_qmmm_0}
 \mathcal{F}[\densqm,\densmm] = \min_{(\hatrdmqm,\rdmmm)\rightarrow (\densqm,\densmm)} f[\hatrdmqm,\rdmmm],
\end{equation}
such that one can rewrite the Gibbs free energy as 
\begin{equation}
 \label{eq:ll_3}
 \potcanoz = \min_{(\densqm,\densmm)} 
\big( (\vextmm|\densmm) + (\vextqm|\densqm)+ \mathcal{F}[\densqm,\densmm] \big).
\end{equation}
\gj{This provides a constructive  proof of the existence of a variational principle for the Helmholtz Free energy, $\potcanoz$, as a functional of one-body densities.  }

\subsection{Explicit coupling in terms of eDFT and cDFT}
\label{sec:qmmm_expl}
\gj{Although Eq. \eqref{eq:ll_3} expresses the Helmholtz free energy only in terms of the one-body densities 
$(\densqm,\densmm)$, the functional itself remains unknown in practice. 
Thus, the explicit connection with well known universal quantum and classical density functionals is still unclear. 
This paragraph focuses on rewriting of $\mathcal{F}[\densqm,\densmm] $ 
so that the functionals of Mermin\cite{Mermin-PR-65}  and Evans\cite{Evans-AP-79} explicitly appear, along with a new QM/MM correlation functional. }

We begin by recalling the universal purely quantum one-body density functionals of Mermin 
\begin{equation}
 \label{eq:mermin}
 \fqm[\densqm] = \min_{\hatrdmqm\rightarrow \densqm } 
 \text{Tr}\{\hatrdmqm\big( \hatboldts  + \hatboldwqm + \kb T  \log(\hatrdmqm)\big)\},
\end{equation}
where $\hatrdmqm$ is a quantum $N$-body density matrix 
and  $(\hatrdmqm\rightarrow \densqm)$ means that the minimization is performed over $N$-body quantum density matrices 
$\hatrdmqm$ leading to a given quantum one-body density $\densqm$. 
Similarly, we can define the universal classical one-body functional of Evans 
\begin{equation}
 \label{eq:evans}
 \fmm[\densmm] = \min_{\rdmmm\rightarrow \densmm} 
 \text{Tr}\{\rdmmm \big( \boldwmm + \kb T \log(\rdmmm)\big)\},
\end{equation}
where $\rdmmm$ is a classical $N$-body probability distribution and 
$(\rdmmm\rightarrow \densmm)$ means that the minimization is performed over $N$-body classical probability distribution 
$\rdmmm$ leading to a given classical one-body density $\densmm$. 
Historically, both Mermin and Evans introduced their universal functionals in the Grand Canonical ensemble, 
so Eqs. \eqref{eq:mermin} and \eqref{eq:evans} are actually the Canonical counter part of the original functionals. 
Also, in their respective original papers\cite{Mermin-PR-65,Evans-AP-79}, neither Mermin or Evans explicitly used the 
constrained search framework which was formalized later by Levy and Lieb\cite{Levy-PNAS-79,Lieb-IJQC-83}.

We can then rewrite $\mathcal{F}[\densqm,\densmm]$ as follows 
\begin{equation}
 \label{eq:ll_4}
 \begin{aligned}
 \mathcal{F}[\densqm,\densmm] =  & \fqm[\densqm] + \fmm[\densmm]  
 + \varepsilon_{\qm}^{\mm}[\densqm,\densmm] + \delta\mathcal{F}_{\qm}^{\mm}[\densqm,\densmm],
 \end{aligned}
\end{equation}
where $\varepsilon_{\qm}^{\mm}[\densqm,\densmm]$ is the mean-field QM/MM interaction 
\begin{equation}
 \begin{aligned}
 \varepsilon_{\qm}^{\mm}[\densqm,\densmm] 
    &=\int\dqss\int\text{d}\qb\wqmmm(\qb,\qs)\densmm(\qb)\densqm(\qs),\\
 \end{aligned}
\end{equation}
which is the QM/MM analogue of the Hartree-term in eDFT, 
and where $\delta\mathcal{F}_{\qm}^{\mm}[\densqm,\densmm]$ 
is the QM/MM correlation functional, \textit{i.e}, the complementary functional 
\begin{equation}
 \label{eq:def_delta_f}
 \delta\mathcal{F}_{\qm}^{\mm}[\densqm,\densmm] = \mathcal{F}[\densqm,\densmm] -\big( 
\fqm[\densqm] + \fmm[\densmm]  + \varepsilon_{\qm}^{\mm}[\densqm,\densmm] \big).
\end{equation}
The functional $\delta\mathcal{F}_{\qm}^{\mm}[\densqm,\densmm]$ in Eq. \eqref{eq:def_delta_f} is there 
to reproduce all correlation terms arising because of the QM/MM interaction. 
The latter contains therefore some energetic and entropic terms. 

We can then rewrite explicitly $\potcanoz$ as 
\begin{equation}
 \begin{aligned}
 \label{eq:ll_5}
 \potcanoz = \min_{(\densqm,\densmm)} 
\big\{ & (\vextqm|\densqm)+ \fqm[\densqm] + \fmm[\densmm] + (\vextmm|\densmm) \\ 
      & + \varepsilon_{\qm}^{\mm}[\densqm,\densmm] + \delta\mathcal{F}_{\qm}^{\mm}[\densqm,\densmm]\big\},
 \end{aligned}
\end{equation}
where we can explicitly recognize the purely quantum and purely classical free-energy functionals, 
together with the QM/MM coupling expressed as the sum of the mean-field and correlation terms. 
If $\delta\mathcal{F}_{\qm}$ is ignored, we recover the mean-field coupling between the eDFT and cDFT functionals that is generally used in the literature\cite{petrosyan_joint_2005,petrosyan_joint_2007,tang_solvent_2020,JeaLevBor-JCTC-20,LabGinJea-JCP-24}.
\manu{We highlight here that, although the outcome of the present derivation is rather natural as it consists of a multi component DFT formalism, 
the explicit definition of $\delta\mathcal{F}_{\qm}^{\mm}[\densqm,\densmm]$ in Eq. \eqref{eq:def_delta_f} 
can help in future developments of approximations for $\delta\mathcal{F}_{\qm}^{\mm}[\densqm,\densmm]$. }
\gj{For instance, the variational formulation in terms of full-body densities allows the use the adiabatic connection to reformulate the problem, and therefore potentially to import several successful strategies used in eDFT to approximate correlation functionals. }

\subsection{Applications to solvation problems}
\label{sec:discuss}

The development presented here were motivated by the necessity of a well grounded theoretical framework for a DFT description of a quantum solute solvated in a classical solvent. This section discusses the application of Eq. \eqref{eq:ll_5} to this particular problem. 

As common in electronic DFT calculations with an implicit solvent description, the nuclei of the molecular solute are considered clamped. The Hamiltonian of the quantum system of Eq. \eqref{eq:hqm} is therefore the Born-Oppenheimer Hamiltonian. 

Regarding now the classical Hamiltonian, it strongly depends on the nature of the considered solvent. We simply assume here that the external part 
of this potential, $\boldvextmm$, represents the interaction between the solvent and the classical nuclei of the solute,
 and that it is local and expressed as a sum of one-body terms. 
Therefore, $\boldvextmm$ naturally contains a Coulombic term but additional terms can be incorporated, in order, for instance, to model repulsion and dispersion interactions. There exists multiple types of classical potential  modeling dispersion and repulsion, the most popular one being the Lennard-Jones potential. 

Central to the description of the QM/MM interaction, $\hatboldwqmmm$, accounts for the electron-solvent interaction. This is a part of the total Hamiltonian that is both unknown and whose quality is critical since it essentially accounts for the information that has been lost when replacing quantum particles by classical particles. This interaction is often modeled using a Coulomb potential. Since only the  external potentials are required to be local to derive Eq. \eqref{eq:ll_5}, more sophisticated models incorporating non-local 
quantum operators can also be incorporated in  $\hatboldwqmmm$. This is a route that should be explored to improve the quality of the QM/MM model.

We now assume that, in the absence of the solvent, the typical electronic gap is large compared to the thermal agitation $\kb T$, 
allowing us to neglect the contribution of electronic entropy in $\fqm[\densqm]$ (see Eq. \eqref{eq:mermin}). 
This is equivalent to considering only the ground state of the electronic quantum system. 
To illustrate the validity of this approximation for systems at room temperature ($T=300\,K$), 
we note that the Boltzmann weight ratio between the ground state and an excited state with a gap of 0.25 eV 
($\approx 0.01\,\text{Ha}$) 
is approximately $10^{-5}$. For a gap of 1.63 eV ($\approx 0.06\,\text{Ha}$), 
which marks the onset of the visible spectrum, this ratio drops dramatically to $10^{-28}$. 
Since most chemical systems near their equilibrium geometry exhibit electronic gaps of several eV, 
this approximation is well-suited for typical room-temperature conditions.
As a consequence, we can take the limit $T \rightarrow 0$ in the definition of the Mermin universal functional 
(see Eq. \eqref{eq:mermin}) to recover the usual universal Levy-Lieb electronic density functional, namely
\begin{equation}
 \fqm[\densqm] \approx \min_{\Psi\rightarrow \densqm } 
  \elemm{\Psi}{\hatboldts  + \hatboldwqm }{\Psi}.
\end{equation}

Regarding solvent effects, we focus on cases where the solvent-solute interaction is not too strong, 
meaning there is no covalent bonding between the solute and solvent species. Under these conditions, 
the previous $T=0~K$ assumption for $\fqm[\densqm]$ remains reasonable, as the electronic structure of the solute 
is not significantly altered by the presence of the solvent.

\gj{Previous attempts to apply  mixed electronic-classical density functional theory to  QM/MM systems were carried on in a semi-Grand canonical ensemble, where the number of classical particles is allowed to fluctuate. This choice was motivated by the availability of classical functionals, which are naturally formulated in the grand canonical ensemble. In the semi-Grand canonical ensemble, the state function is the grand potential, $\Omega_0$, for which the following variational principle, analogous to Eq.~\eqref{eq:ll_5},  stands}

\begin{equation}
 \begin{aligned}
 \label{eq:solv_omega}
 \Omega_0 = &\min_{(\densqm,\densmm)} \big\{   \fqm[\densqm] + \Omega_\mm[\densmm^{\text{GC}}] + \delta \Omega_{\qm}^\mm [\densqm, \densmm^{\text{GC}}] \\
 &+ \varepsilon_{\qm}^\mm [\densqm, \densmm^{\text{GC}}] + \left( \vextqm \mid \densqm \right) + \left( \vextmm \mid \densmm^{\text{GC}} \right)\big\}.
 \end{aligned}
\end{equation}
\gj{In Eq.~\eqref{eq:solv_omega}, $\densmm^\text{GC}$ is the one-body density of classical particles, $\Omega_\mm [\densmm^{\text{GC}}]$ is the grand canonical universal functional \textit{a la} Evans,   $\vextmm$ is the external classical potential, $(\vneqm|\densqm)+ \fqm[\densqm]$) is  the usual quantum Born-Oppenheimer electronic functional and $\delta \Omega_{\qm}^\mm [\densqm, \densmm^{\text{GC}}]$ is the unknown QM/MM correlation term. Eq.~\eqref{eq:solv_omega} can be obtained following the same route that was used to derive Eq.~\eqref{eq:ll_5}. Its explicit derivation is in available in S.I.~III.} \\
 \gj{Minimizing  Eq.~\eqref{eq:solv_omega} in practice require some approximations. First, the choice of the functionals describing  the quantum system and the liquid considerably influences the quality of the predictions. However, determining the appropriate electronic functional for a given solute and the suitable functional for a classical fluid are complex topics that themselves deserve a thorough review. Therefore, we let this discussion aside to focus on the QM/MM coupling and the several strategies adopted in the litterature.\\
In Petrosyan \textit{et al}'s joint DFT \cite{petrosyan_joint_2005,petrosyan_joint_2007}, the interaction of the solvent with both the solute nuclei and electrons is dealt as a whole and the coupling is described by the use of pseudo potential. Thus Eq.~\eqref{eq:solv_omega} solely contains the quantum functional $\fqm$, the solvent one $\fmm$ and  $\varepsilon_{\qm}^{\mm}$. The coupling term is modeled as a dielectric screening due to a non local dielectric function depending on the electron density and a mean-field term between both densities interacting through a Gaussian repulsive potential. }\\
In Tang \textit{et al}\cite{tang_solvent_2020} Reaction DFT, both electronic and classical densities are not optimized simultaneously. Instead, the quantum functional is optimized in a continuum model. The ground state electronic density is used to compute point charges on the solute atoms which generate an external electrostatic potential in the subsequent classical functional optimization. The interaction between the quantum and the classical parts is similar to Petrosyan's work, \textit{i.e} the contribution of the electron and nuclei are gathered together. It is limited to the mean-field term, here a Coulombic term between point charges and a Lennard-Jones term and the correlation functional $ \delta\mathcal{F}_{\qm}^{\mm}$ is lacking. Moreover, since both density are not optimized self-consistently the interaction term $\varepsilon_{\qm}^{\mm}$ only influences the classical optimization and not the quantum one.

In our previous works\cite{JeaLevBor-JCTC-20,LabGinJea-JCP-24}, we also ignored the correlation functional  $ \delta\mathcal{F}_{\qm}^{\mm}$ in Eq. \eqref{eq:ll_5} and limit ourselves to the mean-field term.  The interaction potential was also the sum of a Lennard-Jones and a Coulombic term. However, we do not resort to point charges representation to compute the electrostatic energy but directly use the electronic density
\begin{equation}
 \begin{aligned}
 \varepsilon_{\qm}^{\mm}[\densqm,\densmm] & = \int d\qs \int dr \frac{\densqm(q)  n_C(r) }{|r-\qs|} 
 \end{aligned}
\end{equation}
where $n_C$ is the solvent charge density. 

We also proposed to couple wave function theory-based methods, such as  Hartree-Fock and selected configuration interaction   
 with cDFT\cite{LabGinJea-JCP-24} 
to approximate the universal Levy-Lieb functional $\fqm[\densqm]$ of Eq. \eqref{eq:ll_5}. 

The results obtained in these works\cite{petrosyan_joint_2005,petrosyan_joint_2007,tang_solvent_2020,JeaLevBor-JCTC-20,LabGinJea-JCP-24} clearly indicate that, for ground state properties 
such as dipoles for instance, the mean-field QM/MM approximation gives already a decent improvement with respect to 
\textit{in vaccuum} pure QM calculations.

\section{Conclusion}
\label{sec:conclu}
In the present work we have developed exact variational formulations of the Helmholtz free energy of a $\qmb$/$\mmb$ system 
made of $\nqm$ and $\nmm$ identical particles. 
An important result is that the exact Helmholtz free energy of a $\qmb$/$\mmb$ system can be obtained though a minimization 
of functionals over objects of reduced dimensions, namely the one-body classical and quantum density.

\manu{The assumptions of these derivations is a QM/MM partitioning of the system, together with models for both the MM part and the QM/MM interaction. 
We therefore do not discuss the QM/MM crossover nor how to go beyond the classical models.} 
The starting point of these derivations is the proposal of the $\qmb$/$\mmb$ equivalent to the quantum ground state variational principle or 
the classical Gibbs measure variational principle (see Sec. \ref{sec:exact_qmm}).
This $\qmb$/$\mmb$ functional is formalized in terms of a $\nmm$ classical probability density and 
an $\nqm$ quantum density matrix which also parametrically depends on the $\nmm$ classical variables 
(see Sec. \ref{sec:f_prop_proof} for proofs of the variational property).  
The latter was proposed thanks to the knowledge of the equilibrium density of the $\qmb$/$\mmb$ system (see Eqs. \eqref{eq:eq_qmmm_rho} and \eqref{eq:eq_qmmm_z}) 
which can be obtained from first principle through a semi classical expansion of a partial Wigner transformation 
(see Ref. \onlinecite{NieKapCic-JCP-01} and the section II of the supplementary information for a detailed derivation). 

Starting from this $\qmb$/$\mmb$ variational principle written in terms of $N$-body objects and thanks to constrained search formalisms 
similar to that proposed by Levy and Lieb in eDFT,  we establish a universal QM/MM functional depending only on the one-body classical and quantum objects (see Sec. \ref{sec:levy-lieb}). Building on these results, we explicitly make a coupling with the finite temperature eDFT and cDFT functionals, which allows us to rigorously link the two theories for the description of a QM/MM system (see Sec. \ref{sec:qmmm_expl}). 
We then propose a practical application of this theory in the case of solvation problems, and propose a mean-field treatment. 
Thanks to these derivations, it becomes clear that the explicit correlation between the classical and quantum electronic systems, were implicitly neglected in our previously reported QM/cDFT calculations\cite{JeaLevBor-JCTC-20,LabGinJea-JCP-24}. 

\section*{Supplementary materials}
 The supplementary materials document contains additional information concerning: 
 I) a recall of the variational formulation of statistical quantum mechanics in the canonical ensemble, II) an alternative derivation of the adiabatic QM-MM equilibrium density which does not rely on the Liouville equations or Poisson bracket operators, and III) an extension of the QM/MM functional to the semi-grand-canonical ensemble where the number of classical particles is allowed to vary.

\appendix
\renewcommand{\theequation}{A.\arabic{equation}}
\setcounter{equation}{0}  
\section*{Appendix A: Properties of the functional $\ecano[\hatrdmqm,\rdmmm]$}
\label{sec:f_prop_proof}
We wish to show that i) the functional $\ecano[\hatrdmqm,\rdmmm] $ of Eq. \eqref{eq:pot_cano} is bounded from below by the 
Helmholtz free energy $\potcanoz = -\kb T\log(\eqz )$ with $\eqz$ defined in Eq. \eqref{eq:eq_qmmm_z}, 
and ii) that the only minimizer of $\ecano[\hatrdmqm,\rdmmm] $ is $\eqrdm$ of in Eq. \eqref{eq:eq_qmmm_rho}. 
We will use Jensen's inequality from convex analysis and probability theory to prove these two statements straightforwardly. 
This path is similar to that proposed by Friedli and Velenik in the case of classical lattice systems (see Lemma 6.74 in Ref. \onlinecite{FriVel-book-17}) 
but here applied to the mixed $\qmb$/$\mmb$ case.  
In order to be able to use Jensen's inequality, we will first need to show that the couple $(\hatrdmqm,\rdmmm)$ is indeed a probability measure. 
While for $\rdmmm$ it is obvious, it requires a little more attention for $\hatrdmqm$ as it consists formally in a operator. 
Then, $\ecano[\hatrdmqm,\rdmmm]$ will be written as an average value on probability measures, 
and Jensen's inequality and strict convexity will do the rest. 

\subsection{Rewriting $\ecano[\hatrdmqm,\rdmmm]$ as an average value on probability measures}
We use the adiabatic basis ${\basisq = \{\ket{\Psi_i(\boldqb)}},0\le i\le \infty\}$ 
to develop the quantum or mixed quantum-classical operators.  
More precisely, $\basisq$ consists in the complete set of eigenvectors of the following Schrodinger equation 
\begin{equation}
 \begin{aligned}
 \big( \hathtotqm + \hatboldwqmmm(\boldqb)\big)\ket{\Psi_i(\boldqb)} = E_i(\boldqb) \ket{\Psi_i(\boldqb)},
 \end{aligned}
\end{equation}
such that the Hamiltonian $\hathqm + \hatboldwqmmm(\boldqb)$ is naturally written as 
\begin{equation}
 \label{eq:def_adiab}
 \hathtotqm + \hatboldwqmmm(\boldqb) = \sum_i E_i(\boldqb)\ket{\Psi_i(\boldqb)}\bra{\Psi_i(\boldqb)},
\end{equation}
and the operator $\hatrdmqm(\boldqb)$ is written as follows 
\begin{equation}
 \hatrdmqm(\boldqb) = \sum_i \rdmqm_{ii}(\boldqb)\ket{\Psi_i(\boldqb)}\bra{\Psi_i(\boldqb)}.
\end{equation}
Because of the properties of density matrices, the real-numbers $\rdmqm_{ii}(\boldqb)$ fulfill 
\begin{equation}
 \label{eq:prop_dens}
 \rdmqm_{ii}(\boldqb) \ge 0, \sum_{i \in \basisq} \rdmqm_{ii}(\boldqb) = 1.
\end{equation}
Therefore, as $\rdmmm(\boldqb)$ is a probability distribution and because of the properties of Eq. \eqref{eq:prop_dens},
if we define the function $\mu(\boldqb,i)$ as 
\begin{equation}
 \begin{aligned}
 &{\mathbb R}^{3\nqm}\times {\mathbb N} \rightarrow {\mathbb R}, \\
 &(\boldqb,i)\mapsto \mu(\boldqb,i) = \rdmmm(\boldqb)\rdmqm_{ii}(\boldqb), 
 \end{aligned}
\end{equation}
then $\mu(\boldqb,i)$ fulfils the requirements for a probability measure, \textit{i.e.}  
\begin{equation}
 \begin{aligned}
 \mu(\boldqb,i) \ge 0\,\,\forall\,\, (\boldqb,i), \quad \intdqb \sum_{i}\mu(\boldqb,i)=1.
 \end{aligned}
\end{equation}
For the sake of compactness, we define $\bigx=(\boldqb,i)$, such that we can rewrite the probability measure as 
\begin{equation}
 \begin{aligned}
 \mu(\boldqb,i) \equiv \mu(\bigx),  \quad \intdqb \sum_{i} \equiv \intbigx.  
 \end{aligned}
\end{equation}
The equilibrium density $\hateqrdm$ of Eq. \eqref{eq:eq_qmmm_rho} can be written using the adiabatic basis of Eq. \eqref{eq:def_adiab} as 
follows 
\begin{equation}
 \label{eq:eq_qmmm_rho_adiab}
 \begin{aligned}
 {\hateqrdm}(\boldqb) & = \frac{1}{\eqz}
 e^{-\beta(\hathqm + \hatboldwqmmm(\boldqb))}e^{-\beta\hathtotmm(\boldqb)} \\
 & = \frac{1}{\eqz} \sum_i e^{-\beta\epsilon(\boldqb,i)} \ket{\Psi_i(\boldqb)} \bra{\Psi_i(\boldqb)},\\
 \end{aligned}
\end{equation}
which corresponds to the measure $\mux=\mueqx $ defined as follows
\begin{equation}
 \mueqx \equiv \frac{e^{-\beta\epsilon(\bigx)}}{\eqz},
\end{equation}
where 
\begin{equation}
 \begin{aligned}
 \epsilon(\bigx) = \epsilon(\boldqb,i) \equiv E_i(\boldqb) + \hathtotmm(\boldqb) .
 \end{aligned}
\end{equation}

We introduce the notation for an average value over a probability measure $\mux$ as 
\begin{equation}
 \intbigx \mux f(\bigx) = \langle f \rangle_\mu,
\end{equation}
such that the functional $\ecano[\hatrdmqm,\rdmmm]$ can be rewritten as an average value on the measure $\mu(\bigx)$
\begin{equation}
 \label{eq:ecano_mu}
 \begin{aligned}
 \ecano[\hatrdmqm,\rdmmm] = & \tr\big\{\hatrdmqm\rdmmm\big( \hathtotmm + \hathqm + \hatboldwqmmm + \kb T \log(\hatrdmqm\rdmmm)\big)\} \\
                          = & \intdqb \sum_{i }\mu(\boldqb,i)\big( \epsilon(\boldqb,i) + 
                          \kb T\log(\mu(\boldqb,i)) \\
                          = & \intbigx\mu(\bigx)\big( \epsilon(\bigx) + \kb T\log(\mu(\bigx)) \\
                          = & -\kb T\intbigx\mu(\bigx)\log(\frac{e^{-\beta\epsilon(\bigx)}}{\mu(\bigx)}) \\
                          = & -\kb T\langle \log(\frac{e^{-\beta\epsilon}}{\mu}) \rangle_{\mu} \\
                          \equiv & \ecano[\mu] .
 \end{aligned}
\end{equation}
According to Eq. \eqref{eq:ecano_mu}, the evaluation $\ecano[\hatrdmqm,\rdmmm]$ at $\hateqrdm$ leads straightforwardly to 
\begin{equation}
 \label{eq:ecano_mu_eq}
 \begin{aligned}
 \ecano[\mueq] =  & -\kb T \log(\eqz), 
 \end{aligned}
\end{equation}
which is of course the correct Helmholtz free energy. 
Then two questions remain: i) is $-\kb T \log(\eqz)$ a global minima and ii) is $\hat{\eqrdm}$ the only minimizer ?

\subsection{Use of Jensen inequality to prove variational character}
As the function $-\log(x)$ is strictly convex, we will now use Jensen's inequality from convex analysis and probability measures 
to prove that $\ecano[\mu]$ is bounded from below by the Helmholtz free energy and that $\mueq$ is the only minimizer. 
The Jensen's inequality can be stated as follows. 

If $\varphi$ is a convex function (\textit{i.e.} always above its tangent lines), 
$\mu(\bigx)$ a probability measure on $\bigx \in E$ where $E$ is a normed vector space,  
$g(\bigx)$ a function from $E$ to ${\mathbb R}$, then Jensen's equality states that
\begin{equation}
 \label{eq:jensen_av}
 \varphi(\langle g \rangle_{\mu}) \le \langle \varphi(g)\rangle_{\mu}, 
\end{equation}
or written in terms of integrals
\begin{equation}
 \label{eq:jensen_int}
 \varphi\bigg( \int \text{d}\bigx \mu(\bigx) g(\bigx)\bigg) \le 
 \int \text{d}\bigx \mu(\bigx) \varphi\big(g(\bigx)\big).
\end{equation}

As $-\log(x)$ is convex, the Jensen's equality can be applied to $\ecano[\mu]$ of Eq. \eqref{eq:ecano_mu} by setting
\begin{equation}
 \begin{aligned}
 &g(\bigx) = \frac{e^{-\beta\epsilon(\bigx)}}{\mux},\\
 &\varphi(x) = -\kb T \log(x).
 \end{aligned}
\end{equation}
Therefore the inequality of Eq. \eqref{eq:jensen_av} becomes 
\begin{equation}
 \label{eq:jensen_f}
-\kb T \log\langle \frac{e^{-\beta\epsilon}}{\mu} \rangle_{\mu} \le -\kb T\langle \log(\frac{e^{-\beta\epsilon}}{\mu}) \rangle_{\mu}, 
\end{equation}
and as 
\begin{equation}
 \begin{aligned}
 \langle \frac{e^{-\beta\epsilon}}{\mu} \rangle_{\mu} & = \intbigx \mux \frac{e^{-\beta\epsilon(x))}}{\mux} \\
                                                      & = \eqz,
 \end{aligned}
\end{equation}
one can rewrite Eq. \eqref{eq:jensen_f} as 
\begin{equation}
 \label{eq:jensen_f2}
 \begin{aligned}
 -\kb T \log(\eqz) \le \ecano[\mu],
 \end{aligned}
\end{equation}
which guarantees the variational character. 
Eventually, the strict convexity of the $-\log(x)$ function implies that the Jensen's inequality becomes an equality 
if and only if $g(\bigx)=cst$, which translates into 
\begin{equation}
 g(\bigx)=cst \Leftrightarrow \frac{e^{-\beta\epsilon(\bigx)}}{\mux} = cst \Leftrightarrow \mux \propto e^{-\beta\epsilon(\bigx)},
\end{equation}
therefore implying that $\mueqx$ is the only minimizer. 
We point out that this proof essentially consists in the Gibb's inequality re-framed with convex analysis vocabulary. 

\bibliography{biblio_main}

 \end{document}


\begin{abstract}
 This document contains additional information concerning: I) a recall of the variational formulation of statistical quantum mechanics in the canonical ensemble, II) an alternative derivation of the adiabatic QM-MM equilibrium density, and III) an extension to the semi-grand canonical ensemble of the DFT formulation for a QM/MM problem. 
\end{abstract}

\title{A variational formulation of the free energy of mixed quantum-classical systems: coupling classical and quantum density functional theories
: supplementary information}
\author{Guillaume Jeanmairet}
\email{guillaume.jeanmairet@sorbonne-universite.fr}
\affiliation{Sorbonne Université, CNRS, Physico-Chimie des Électrolytes et Nanosystèmes
Interfaciaux, PHENIX, F-75005 Paris, France}
\affiliation{Réseau sur le Stockage Électrochimique de l'Énergie (RS2E), FR CNRS
3459, 80039 Amiens Cedex, France}
\author{Maxime Labat}
\affiliation{Sorbonne Université, CNRS, Physico-Chimie des Électrolytes et Nanosystèmes
Interfaciaux, PHENIX, F-75005 Paris, France}
\affiliation{Réseau sur le Stockage Électrochimique de l'Énergie (RS2E), FR CNRS
3459, 80039 Amiens Cedex, France}
\author{Emmanuel Giner$^*$}%
\email{emmanuel.giner@lct.jussieu.fr}
\affiliation{Laboratoire de Chimie Théorique, Sorbonne Université and CNRS, F-75005 Paris, France}

\maketitle
\section{Recall of the variational formulation of statistical quantum mechanics in the canonical ensemble}
\label{eq:var_phistat}
Just as the ground state energy and wave function is often formalized in terms of the variational principle, 
there exists a similar functional formulation of quantum statistical mechanics. 
As it might not be widely spread in the quantum chemistry community, which consists in the main audience of this paper, 
we recall here this variational formulation. 

Let us consider a $\qmb$ system composed of $N$ particles, not necessarily identical, described by the Hamiltonian $\fulhn$. 
In the canonical ensemble, the equilibrium density (ED) matrix $\rdmneq$ has the following expression
\begin{equation}
 \label{eq:def_ed_cano}
 \rdmneq = \frac{e^{-\beta \fulhn}}{\zfunc},
\end{equation}
with $\beta^{-1} = k_B T$ and $\zfunc$ is the partition function  
\begin{equation}
 \zfunc = \trn\{ e^{-\beta \fulhn} \},
\end{equation}
where we introduced the trace of an $N$-electron operator, $\hat{O}^N$, which is given in real space by 
\begin{equation}
 \begin{aligned}
  \trn\{ \hat{O}^N \} & = \int \text{d}q_1 \hdots \text{d}q_{N} \elemm{\rlots{N}}{\hat{O}}{\rlots{N}} \\
                    & = \int \text{d}q_1 \hdots \text{d}q_{N} \oprr. 
 \end{aligned}
\end{equation}
The corresponding Helmholtz free energy is then obtained by
\begin{equation}
 \label{eq:def_pot_cano}
 \potcanoz = - \frac{1}{\beta} \log(\zfunc).
\end{equation}
Any other thermodynamic equilibrium properties  can be obtained from the knowledge of $\rdmneq$ using ensemble average. 

Although those well established results can be derived directly by arguments inherent to statistical mechanics, 
they can also be obtained through a variational formulation of statistical mechanics, 
which was first proposed by Gibbs in the classical case and latter on extended to quantum mechanics by Von Neumann\cite{Neumann-Pri-96}.
In the latter framework, $\potcanoz$ can be expressed as the minimum of a free-energy density matrix functional 
\begin{equation}
 \label{eq:pot_cano_1}
 \potcanoz = \min_{\rdmn} \potcano[\rdmn], 
\end{equation}
where 
$\rdmn$ are $N$-representable density matrices, \textit{i.e.} density matrices representing statistical mixing of $N$-particles density matrices.
$\potcano[\rdmn]$ is the free-energy functional associated to the density matrix $\rdmn$,
\begin{equation}
 \label{def:potcano}
 \potcano[\rdmn] = {E[\rdmn] - \frac{1}{\beta}{S}[\rdmn]}, 
\end{equation}
{where $E[\rdmn]$ is the energy functional associated to $\rdmn$
\begin{equation}
E[\rdmn] = \tr^N \{\rdmn \fulhn \},
\end{equation}}
and $S[\rdmn]$ is the dimension-less entropy functional associated to $\rdmn$ 
\begin{equation}
 {S}[\rdmn] = -\tr^N \{\rdmn \log(\rdmn)\}.
\end{equation}
The density matrix minimizing Eq.\eqref{eq:pot_cano_1} is then the ED matrix $\rdmneq$ of Eq. \eqref{eq:def_ed_cano}, 
and the functional of Eq. \eqref{def:potcano} evaluated at $\rdmneq$ is the Helmholtz free energy $\potcanoz$ 
of Eq. \eqref{eq:def_pot_cano}. 

Two special cases of this variational framework are worth mentioning in the present context. 
Firstly, at zero temperature, the entropic term vanishes and the corresponding ED matrix is nothing 
but $\rdmneq = \ket{\Psi_0}\boldqsa{\Psi_0} $, where $\ket{\Psi_0}$ is the ground state of $\fulhn$. 
This is then equivalent to the variational principle routinely used in quantum chemistry, 
but here it is formalized in terms of $N$-body density matrices rather than $N$-body wave functions. 
Secondly, in the classical case, the density matrices $\rdmn$ reduce to probability distributions in terms of 
momentum/position variables, and the Hamiltonian operator becomes the classical Hamiltonian function. 
The latter can be obtained from first principle through an expansion in powers of $\hbar$ of the Wigner transformation 
of the fully quantum system, and then keeping the zeroth-order term, 
as shown by Wigner in his seminal work on semi-classical expansion\cite{Wigner-PR-32}. 
The variational formulation reduces then to that proposed by Gibbs, \textit{i.e.}  
the Helmholtz free energy is the minimum of the classical free energy functional analog of Eq. \eqref{def:potcano}, 
that is reached for the equilibrium probability distribution.

The advantage of a variational framework is that one can apply all the tools from functional analysis 
to reformulate this cumbersome problem in simpler ways. 
Also, alternative formulations can be obtained by reducing the dimensionality of the functions 
on which one performs the minimization. 
Striking examples of the advantage of the latter are certainly 
the universal Levy-Lieb functional of the electron density of eDFT at zero temperature, 
or the formulation of the classical statistical problems of cDFT.  

\section{Alternative derivation of the ED for a $\qmb$/$\mmb$ system}
\label{sec:deriv_ed_qmmm}
\subsection{Big picture and preliminaries}
In this section, we report a derivation of the ED for a mixed $\qmb$/$\mmb$ system that does not rely on 
the usual Bloch equations, Livouillian operator and Poisson bracket operator familiar to the quantum dynamics community\cite{NieKapCic-JCP-01}, 
but here we rather follow a path similar to Wigner\cite{Wigner-PR-32} which, we believe, is easier to follow for 
the quantum chemistry community. We nevertheless obtain the same result of Nielsen \textit{et. al.}\cite{NieKapCic-JCP-01}. 

The big picture is the following: we start from a fully quantum system containing $\nqm+\nmm$ particles, 
and we consider the ED in the canonical ensemble. 
Then, we make a semi-classical expansion of the Wigner transformation of the ED, where only the 
$\nmm$ variables corresponding to the heavy particles have been transformed.  
Eventually, we make a semi classical expansion in terms of the ratio of the masses $m/M$ 
and we keep only the zeroth-order term in order to recover purely classical particles for the $\mmb$ part. 

Therefore, we start with a fully quantum system where the $\nmm$ particles, 
that will eventually become classical after the semi-classical expansion, are now treated with the 
quantum operator 
\begin{equation}
 \hathmm = \hatboldtb + \hatboldvextmm + \hatboldwmm, 
\end{equation}
with 
\begin{equation}
 \hatboldtb = \alpha^2\sum_{i=1}^{\nmm}\frac{\hatpb{i}^2}{2 m} , 
 \qquad \hatpb{i} =  -i\hbar \nabla_{\varqb{i}}, \qquad 
\end{equation}
where $\alpha = (m/M)^{1/2}$, and 
\begin{equation}
  \hatboldvextmm = \sum_{i=1}^{\nmm} \vextmm(\varqb{i}), \quad \hatboldwmm=\sum_{i>j} \wmm(\varqb{i},\varqb{j}). 
\end{equation}
{We also introduce the following notations for classical vectors and vectors of operators}
{
\begin{equation}
 \begin{aligned}
 &{\arrhatpb} = \sum_{i=1}^{\nmm} \hatpb{i}, 
 \qquad {\arrhatpb}^2 = \sum_{i=1}^{\nmm} (\hatpb{i})^2,\\ 
 &{\arrpb} = \sum_{i=1}^{\nmm} \pb_{i}, 
 \qquad {\arrpb}^2 = \sum_{i=1}^{\nmm} (\pb_{i})^2,\\ 
& {\vec\hatboldqb} = \sum_{i=1}^{\nmm} \hatqb{i}, \qquad 
 {\vec\boldqb} = \sum_{i=1}^{\nmm} \qb_{i}. 
 \end{aligned}
\end{equation}
}
The Hamiltonian of the total system is then fully quantum 
\begin{equation}
 \fulh = \hathqm + \hatboldwqmmm + \hathmm,
\end{equation}
and our goal is then to obtain a semi-classical expansion of the equilibrium density of the system which is, up to a normalization constant, given by 
\begin{equation}
 \begin{aligned}
 \hateqrdm & = e^{-\beta \fulh}.
 \end{aligned}
\end{equation}
The real-space representation of $\hateqrdm$ is then
\begin{equation}
 \label{eq:eqrdm_real}
 \eqrdm(\boldqb ,\boldqs;\boldqpb,\boldqps) = \elemm{\boldqb ,\boldqs}{\hateqrdm}{\boldqpb,\boldqs},
\end{equation}
where, as we assume that $\nqm$ and $\nmm$ particles are not identical, 
the basis for the whole system is expressed as a tensor product of the individual basis, \textit{i.e.} 
\begin{equation}
 \ket{\boldqb ,\boldqs} \equiv \ket{\boldqb} \otimes \ket{\boldqs},
\end{equation}
{
and where we used the notations for the eigenvectors of the position operators 
\begin{equation}
 \begin{aligned}
 &\ket{\boldqb} = \ket{\varqb{1},\hdots,\varqb{\nmm}}, \qquad 
 {\vec\hatboldqb} \ket{\boldqb} = {\arrqb} \ket{\boldqb},\\ 
 &\ket{\boldqs} = \ket{\varqs{1},\hdots,\varqs{\nqm}}, \qquad\,\,\,\,\,\,\,
 {\vec\hatboldqs} \ket{\boldqs} = {\arrqs} \ket{\boldqs}.\\ 
 \end{aligned}
\end{equation}
}

As our goal is to eventually obtain $\nmm$ purely classical particles which are naturally 
described in phase space, we will first exchange the couple of real-space variables $(\boldqb ,\boldqpb)$ 
by a couple of phase-space variables $(\boldqb ,\boldpb )$ in Eq. \eqref{eq:eqrdm_real}, which can be done by a Wigner transformation\cite{Wigner-PR-32}.  
A crucial property of the Wigner transformation is that it allows to obtain traces of quantum operators as classical traces 
in phase space (for a pedagogical introduction to Wigner transform, the reader might refer to Ref. \onlinecite{Case-AJP-08}). 
More precisely, if $\hat{A}$, $\hat{B}$ are two quantum operators, and $\tilde{A}_W(q,p)$,  $\tilde{B}_W(q,p)$ their 
respective Wigner transformation, then the following property holds
\begin{equation}
 \tr\{ \hat{A}\hat{B} \} = \int \text{d}q\text{d}p\tilde{A}_W(q,p)\tilde{B}_W(q,p). 
\end{equation}

In our case, the Wigner transformation of $\hateqrdm$ is performed only on the $\nmm$ variables, which, by the very definition 
of the Wigner transformation, leads to 
\begin{equation}
 \begin{aligned}
 \label{eq:def_wigner}
 \tilde{W}_0(\boldqb ,\boldqs;\boldpb ,\boldqps) &= 
 \int d\boldqpb e^{i{\vec\boldpb} \cdot {\vec\boldqpb}/\hbar} 
 \eqrdm(\boldqb+\frac{\boldqpb}{2},\boldqs;\boldqb-\frac{\boldqpb}{2},\boldqps) \\
 &= \int d\boldqpb e^{i{\vec\boldpb} \cdot \vec{\boldqpb}/\hbar} 
  \elemm{\boldqb+\frac{\boldqpb}{2},\boldqs}{\hateqrdm}{\boldqb-\frac{\boldqpb}{2},\boldqps}
 \end{aligned}
\end{equation}
where here we emphasize that $\boldqb$, $\boldqpb$ and $\boldpb$ are scalars and not operators, 
{and where we introduce the notations 
\begin{equation}
 \begin{aligned}
 & g(\boldqb+\boldqpb) = g(\qb_1+\qb'_1,\hdots,\qb^{\nmm}+\qb'_{\nmm}) \\
 & \ket{\boldqb+\boldqpb} = \ket{\qb_1+\qb'_1,\hdots,\qb^{\nmm}+\qb'_{\nmm}}.
 \end{aligned}
\end{equation}
}

\subsection{Wigner transformation: first steps}
We will now carry some calculations similar to Eqs. (15-16) of the seminal paper of Wigner\cite{Wigner-PR-32}, 
but applied to a mixed quantum-classical system. 
One can rewrite Eq.\eqref{eq:def_wigner} as 
\begin{equation}
 \label{eq:def_wigner_2}
 \begin{aligned}
 &\tilde{W}_0(\boldqb ,\boldqs;\boldpb ,\boldqps)  
 = \int d\boldqpb \\
& e^{i\arrpb \cdot (\arrqb+\frac{\arrpb}{2})/\hbar} 
 \eqrdm(\boldqb+\frac{\boldqpb}{2},\boldqs;\boldqb-\frac{\boldqpb}{2},\boldqps) 
 e^{-i\arrpb \cdot (\arrqb-\frac{\arrqpb}{2})/\hbar} \\
 \end{aligned}
\end{equation}
and by noticing that  
$f(\hatqbz)\ket{\qb'} = f(\qb')\ket{\qb'}  $ 
for any function $F [\eqrdmqm, \eqrdmmm]$, then one can write 
\begin{equation}
 \begin{aligned}
   \ket{\boldqb-\frac{\boldqpb}{2}} e^{-i\arrpb \cdot (\arrqb-\frac{\arrpb}{2})/\hbar} 
& = 
 e^{-i\arrpb \cdot \arrhatqb/\hbar} \ket{\boldqb-\frac{\boldqpb}{2}} ,
 \end{aligned}
\end{equation}
such that we can rewrite the argument of the integral in Eq.\eqref{eq:def_wigner_2} as
\begin{equation}
 \label{eq:wigner_3}
 \begin{aligned}
& e^{i\arrpb \cdot (\arrqb+\frac{\arrqpb}{2})/\hbar} 
 \elemm{\boldqb+\frac{\boldqpb}{2},\boldqs}{\hateqrdm}{\boldqb-\frac{\boldqpb}{2},\boldqps} 
 e^{-i\arrpb \cdot (\arrqb-\frac{\arrqpb}{2})/\hbar} \\
& = 
 \elemm{\boldqb+\frac{\boldqpb}{2},\boldqs}{
 e^{i\arrpb \cdot \hatarrqb /\hbar} \hateqrdm e^{-i\arrpb \cdot \hatarrqb/\hbar} 
 }{\boldqb-\frac{\boldqpb}{2},\boldqps}. \\
 \end{aligned}
\end{equation}
We then see that the argument of the integral in Eq. \eqref{eq:def_wigner_2} is nothing but the off-diagonal matrix element 
of the similarity transformation of $\hateqrdm$ by the operator $\exp({-i\boldpb  \cdot \hatboldqb/\hbar})$. 
Instead of performing the similarity transformation directly on $\exp(-\beta \fulh)$, we use 
\begin{equation}
 \begin{aligned}
  \hat{U} e^{\hat{A}} \hat{U}^{-1} 
 & = e^{\hat{U}\hat{A}\hat{U}^{-1}},
 \end{aligned}
\end{equation}
and first perform the similarity transformation of $\fulh$ and then take the exponential of the similarity transformed Hamiltonian. 
On has then to compute 
\begin{equation}
 \label{eq:h_similarity_1}
 \begin{aligned}
  \wignerh & = e^{i\arrpb \cdot \hatarrqb /\hbar} \fulh e^{-i\arrpb \cdot \hatarrqb/\hbar}, \\
 \end{aligned}
\end{equation}
which is an operator explicitly depending on the parameter $\boldpb$. 
We use the then the Baker-Campbell-Hausdorff relation  
\begin{equation}
 \begin{aligned}
  \wignerh & = \fulh +i \frac{\arrpb }{\hbar} \cdot \big[ \hatarrqb, \fulh \big] - \frac{1}{2}
                    \frac{{\vec\boldpb^2} }{\hbar^2} \big[ \hatarrqb, \big[ \hatarrqb, \fulh \big] \big] + \hdots
 \end{aligned}
\end{equation}
which is essentially an infinite series of nested commutator of $\hatarrqb$ with $\big[ \hatarrqb, \fulh \big]$. 
As $\big[\hatarrqb,\hathqm\big]=0$,
the first commutator yields 
\begin{equation}
 \begin{aligned}
 \big[ \hatarrqb, \fulh \big]& = \big[ \hatarrqb, \frac{\hatarrpb^2}{2M} \big] \\
 & = \frac{i \hbar}{M} \hatarrpb.
 \end{aligned}
\end{equation}
Then the second commutator reads 
\begin{equation}
 \begin{aligned}
 \big[ \hatarrqb, \big[ \hatarrqb, \fulh \big] \big]& =  \frac{i \hbar}{M}\big[ \hatarrqb, \hatarrpb\big] \\
 & = -\frac{\hbar^2}{M}.
 \end{aligned}
\end{equation}
As $\big[ \hatarrqb, \big[ \hatarrqb, \fulh \big] \big]$ yields a constant, all higher order nested commutators 
are zero such that higher order terms vanish. 
Therefore, the similarity transformed Hamiltonian of Eq. \eqref{eq:h_similarity_1} yields 
\begin{equation}
 \begin{aligned}
  \wignerh & = \fulh +i \frac{\arrpb }{\hbar} \cdot \big[ \hatarrqb, \fulh \big] - \frac{1}{2}
                     \frac{\arrpb^2 }{\hbar^2} \big[ \hatarrqb, \big[ \hatarrqb, \fulh \big] \big]  \\
 & = \fulh -\frac{\arrpb }{M} \cdot \hatarrpb 
 + \frac{\arrpb^2 }{2 M}. 
 \end{aligned}
\end{equation}
This is essentially the same result that Wigner obtained in equation 16 of his original paper\cite{Wigner-PR-32} but in the case of a mixed quantum-classical system. 
We note in passing that if we perform the Wigner transformation on $\fulh$ we obtain 
\begin{equation}
 \begin{aligned}
\wignerh(\boldqb ,\boldpb ) =  
& \int d\boldqpb 
\elemm{\boldqb+\frac{\boldqpb}{2},\boldqs}{\wignerh}{\boldqb-\frac{\boldqpb}{2},\boldqps} \\
 \end{aligned}
\end{equation}
but as 
\begin{equation}
 \elemm{\boldqb+\frac{\boldqpb}{2},\boldqs}
{ \hatarrpb^n}
{\boldqb-\frac{\boldqpb}{2}} = (-i\hbar)^n\delta(\boldqpb)^{(n)},
\end{equation}
where $\delta(\boldqpb)^{(n)}$ is the $n$-th order derivative of the delta distribution, 
we then obtain
\begin{equation}
 \begin{aligned}
\wignerh(\boldqb ,\boldpb ) =  \hathqm +  \boldvextmm(\boldqb )  + \boldwmm(\boldqb ) 
+ \frac{\arrpb^2 }{2 M}.
 \end{aligned}
\end{equation}
Therefore as expected, the partial Wigner transformation on the $(\boldqb ,\boldqpb)$ variables 
of the quantum Hamiltonian is the classical function associated with the $(\boldqb ,\boldpb )$ variables. 

\subsection{Classical limit: heavy mass}
\manu{
In order to perform a semi-classical limit and obtain classical heavy particles, we will perform a $m\rightarrow 0$ (or equivalently $M\rightarrow \infty$), which will be done using the scaled variable technique proposed by Kapral and Ciccoti\cite{KapCic-JCP-99}. 
The latter consists in expressing all variables of our problem (\textit{i.e.} $\boldqb ,\boldqs$ and $\boldpb$) 
in terms of the typical energy scale of our problem which is $\epsilon_0 = \beta^{-1}$. 
More precisely, we introduce the thermal wave length of the light particle $\lambda_m = \hbar/(m\epsilon_0)^{1/2}$ 
and the thermal momentum of the heavy particle $P_M = (M \epsilon_0)^{1/2}$. 
We can then re-express $\wignerh$ in terms of the dimension less length variables $\boldtildeqs = \boldqs/\lambda_m$ and $\boldtildeqb = \boldqb/\lambda_m$, 
together with the dimension less momentum variables $\boldtildepb=\boldpb/P_M$. 
The explicit expression in terms of the dimension-less variables of $\wignerh$ reads then 
\begin{equation}
 \label{eq:h_similarity_2}
 \begin{aligned}
  \wignerh = \epsilon_0 \big(  \boldtildepb^2 +\hatvtot(\boldtildeqs,\boldtildeqb)
           - \frac{1}{2}\Delta_{\boldtildeqs} -i\alpha \boldtildepb\cdot \nabla_{\boldtildeqb}  - \alpha^2 \frac{1}{2}\Delta_{\boldtildeqb} \big),
 \end{aligned}
\end{equation}
where $\hatvtot(\boldtildeqs,\boldtildeqb)$ is the dimension less total potential 
\begin{equation}
 \hatvtot(\boldtildeqs,\boldtildeqb) = \frac{1}{\epsilon_0}\big(\hatboldwqm(\boldtildeqs) + \hatboldwmm(\boldtildeqb ) 
        + \hatboldwqmmm(\boldtildeqs,\boldtildeqb ) + \hatboldvextqm(\boldtildeqs) + \hatboldvextmm(\boldtildeqb )\big),
\end{equation}
and we recall that $\alpha = (m/M)^{1/2}$ is the parameter representing the ratio of masses. 
As $\alpha$ represents the parameter that will tend to zero, it is useful to rewrite $\wignerh$ in powers of $\alpha$
\begin{equation}
 \label{eq:h_similarity_3}
 \begin{aligned}
 \wignerh = 
 \wignerhz + \alpha \wignerhone + \alpha^2 \wignerhtwo, 
 \end{aligned}
\end{equation}
with 
\begin{equation}
 \label{eq:h_similarity_6}
 \begin{aligned}
 \wignerhz = \epsilon_0 \big( \boldtildepb^2 - \frac{1}{2}\Delta_{\boldtildeqs}+\hatvtot(\boldtildeqs,\boldtildeqb) \big),
 \end{aligned}
\end{equation}
\begin{equation}
 \label{eq:h_similarity_7}
 \begin{aligned}
 \wignerhone = -i  \epsilon_0 \boldtildepb\cdot \nabla_{\boldtildeqb},
 \end{aligned}
\end{equation}
\begin{equation}
 \label{eq:h_similarity_8}
 \begin{aligned}
 \wignerhtwo = - \epsilon_0 \frac{1}{2}\Delta_{\boldtildeqb},
 \end{aligned}
\end{equation}
such that we can take the limit of $\wignerh$ when $\alpha \rightarrow 0$ 
\begin{equation}
 \label{eq:h_similarity_5}
 \begin{aligned}
 \lim_{\alpha \rightarrow 0}\wignerh  = \epsilon_0 \wignerhz.  
 \end{aligned}
\end{equation}
Coming back to unscaled units, if we define 
\begin{equation}
 \hathtotmm = \hatboldwmm + \hatboldvextmm,
\end{equation}
\begin{equation}
\hathtotqm = \hathqm + \hatboldvextqm,
\end{equation}
the zeroth-order operator $\wignerhz$ can then be written as 
\begin{equation}
 \wignerhz = \hathtotqm + \hatboldwqmmm + \hathtotmm +  \frac{\arrpb^2 }{2M},
\end{equation}
where now appears explicitly the \textit{classical} kinetic energy  of the heavy particles ${\arrpb^2 }/{2M}$, 
but the potential term $\hathtotmm$ remains an operator. 
}

\subsection{Finishing the Wigner expansion}
In order to compute the Wigner transformation of the semi classical ED we have to compute 
\begin{equation}
 \label{eq:def_wigner_3}
 \begin{aligned}
 &\tilde{W}_0(\boldqb ,\boldqs;\boldpb ,\boldqps)  
  = \int d\boldqpb 
 \elemm{\boldqb+\frac{\boldqpb}{2},\boldqs}{\hatwignerrdm^{(0)} }{\boldqb-\frac{\boldqpb}{2},\boldqps}, 
 \end{aligned}
\end{equation}
with 
\begin{equation}
 \label{eq:def_part_wig}
 \hatwignerrdm^{(0)} = e^{-\beta\wignerhz}.
\end{equation}
Because $\boldpb$ in $\wignerhz$ is a function and that $\hathtotmm$ commutes with both 
$\hathqm$ and $\hatboldwqmmm$, we can write the operator $\hatwignerrdm^{(0)}$ as 
\begin{equation}
 \begin{aligned}
  \hatwignerrdm^{(0)}  & = e^{-\beta\big(\hathtotqm + \hatboldwqmmm \big)}e^{-\beta\big(\frac{\arrpb^2 }{2M}  + \hathtotmm\big) }  . \\
 \end{aligned}
\end{equation}
Before computing explicitly the matrix element appearing in Eq. \eqref{eq:def_wigner_3}, 
we compute first the action of $\hatwignerrdm^{(0)}$  on a given basis function $\ket{\boldqb ,\boldqs}$ 
and, as $\hathtotmm$ is diagonal on the $\ket{\boldqb}$ basis, one gets
\begin{equation}
 \label{eq:def_wigner_5}
 \begin{aligned}
 & \hatwignerrdm^{(0)}\ket{\boldqb+\frac{\boldqpb}{2},\boldqs} \\
 & =  {e^{-\beta\big(\hathtotqm + \hatboldwqmmm \big)}} 
 \ket{\boldqb+\frac{\boldqpb}{2},\boldqs}e^{-\beta\big(\frac{\arrpb^2 }{2M}  + \htotmm(\boldqb+\frac{\boldqpb}{2})\big) },\\
 \end{aligned}
\end{equation}
where now $\htotmm$ is a function an not an operator. 
Therefore, one can rewrite Eq. \eqref{eq:def_wigner_5} as 
\begin{equation}
 \label{eq:def_wigner_6}
 \begin{aligned}
 & \hatwignerrdm^{(0)}\ket{\boldqb+\frac{\boldqpb}{2},\boldqs} \\
 & =  {e^{-\beta\big(\hathqm + \hatboldwqmmm \big)}} 
\ket{\boldqb+\frac{\boldqpb}{2},\boldqs}
e^{-\beta\hmm(\boldqb+\frac{\boldqpb}{2},\boldpb ) },\\
 \end{aligned}
\end{equation}
where now the fully classical Hamiltonian function $\hmm(\boldqb ,\boldpb )$ appears. 
Also, the operator $\hathtotqm + \hatboldwqmmm$ is diagonal on the $\ket{\boldqb}$ basis,  
and therefore, the exponential of $\hathqm + \hatboldwqmmm$ is also diagonal on that basis, such that 
\begin{equation}
 \begin{aligned}
& {e^{-\beta\big(\hathtotqm + \hatboldwqmmm \big)}}\ket{\boldqb+\frac{\boldqpb}{2},\boldqs} = \\
&\ket{\boldqb+\frac{\boldqpb}{2}}{e^{-\beta\big(\hathtotqm + \hatboldwqmmm(\hatboldqs,\boldqb+\frac{\boldqpb}{2}) \big)}} \ket{\boldqs}.
 \end{aligned}
\end{equation}
where now $\hatboldwqmmm(\hatboldqs,\boldqb )$ is an operator only on the $\hatboldqs$ variable which depends parametrically on the $\boldqb$ variables (as the Born-Oppenheimer Hamiltonian). 
As a consequence we can write the matrix elements as 
\begin{equation}
 \begin{aligned}
  &\elemm{\boldqb+\frac{\boldqpb}{2},\boldqs}{e^{-\beta\big(\hathtotqm + \hatboldwqmmm \big)}}{\boldqb-\frac{\boldqpb}{2},\boldqps} \\
 =& \elemm{\boldqs}{e^{-\beta\big(\hathtotqm + \hatboldwqmmm(\hatboldqs,\boldqb+\frac{\boldqpb}{2}) \big)}}{\boldqps}
 \boldqsaket{\boldqb+\frac{\boldqpb}{2}}{\boldqb-\frac{\boldqpb}{2}} \\
 =& \elemm{\boldqs}{e^{-\beta\big(\hathtotqm + \hatboldwqmmm(\hatboldqs,\boldqb+\frac{\boldqpb}{2}) \big)}}{\boldqps}
 \delta(\boldqpb).
 \end{aligned}
\end{equation}
We can now compute easily the Wigner transform as in Eq. \eqref{eq:def_wigner_3}
\begin{equation}
 \label{eq:def_wigner_7}
 \begin{aligned}
  \tilde{W}_0(\boldqb ,\boldqs;\boldpb ,\boldqps)  
  = \int d\boldqpb 
 & \elemm{\boldqs}{e^{-\beta\big(\hathtotqm + \hatboldwqmmm(\hatboldqs,\boldqb+\frac{\boldqpb}{2}) \big)}}{\boldqps} \\
 & e^{-\beta\hmm(\boldqb+\frac{\boldqpb}{2},\boldpb ) }\delta(\boldqpb)
 \end{aligned}
\end{equation}
and once integrated it yields 
\begin{equation}
 \label{eq:def_wigner_7}
 \begin{aligned}
 \tilde{W}_0(\boldqb ,\boldqs;\boldpb ,\boldqps) =  
 & \elemm{\boldqs}{e^{-\beta\big(\hathtotqm  + \hatboldwqmmm(\hatboldqs,\boldqb ) \big)}}{\boldqps} e^{-\beta\hmm(\boldqb ,\boldpb )}.
 \end{aligned}
\end{equation}
We obtain then, up to a normalization constant, at the result of Eq. (9) of Sec. II. B of the main text.

\newpage
\section{Extension of the QM/MM functional in a semi-grand-canonical ensemble}
\label{sec:deriv_ed_qmmm}

In this appendix, we aim to establish a rigorous variational formulation of the mixed quantum-classical system in a semi-grand-canonical ensemble. One of the subtleties of this new ensemble is that the number $\nqm$ of quantum particles in the system remains fixed, while the number $\nmm$ of classical particles is allowed to fluctuate. Therefore, the considered semi-grand-canonical ensemble is defined by $\nqm$, $\mu$, $V$, and $T$. \gj{This is the natural ensemble to propose an eDFT/cDFT formalism since classical DFT is usually formulated in the grand canonical ensemble~\cite{Mermin-PR-65,Evans-AP-79}.
We follow a similar route to the one employed for the canonical ensemble reported in the main paper.}
First, we introduce the semi-grand canonical ensemble in Sec. \ref{subsec:gc_form} which allows us to express exactly the QM/MM $N$-body equilibrium density (ED) matrix in Sec.~\ref{subsec:EDM_GC}. Then, in Sec.~\ref{subsec:FormVar}, \manu{noticing that the ED is a special case of more general density matrices,} we develop a variational and exact formulation of the grand potential of the system  as a function of the ED matrix. Lastly, we use the Levy-Lieb constrained-search formalism to rewrite the grand potential as a functional minimization over one-body QM and MM densities in Sec.~\ref{subsec:LevyLieb}

\subsection{Grand canonical formalism with direct sums}
\label{subsec:gc_form}
As the number of classical particles vary, we need to define the meaning of the
grand canonical operators and their associated trace operator. For readers who are more familiar with quantum mechanics, this is equivalent to constructing the Fock space from the Hilbert space. However, in the classical case, the canonical ensemble  is a phase space and the grand canonical ensemble is the disjoint union of all these $\nmm$ particles canonical ensembles. This means that any quantity $A^{\gc}$ in the grand canonical ensemble is defined sectorwise on the phase spaces corresponding to each number of particles $\nmm$. By analogy with the quantum case, we chose to note this disjoint union by using the direct sum symbol, while acknowledging that this is mathematically incorrect.

As we will manipulate quantities with varying number of classical particles, 
we explicitly write $\nmm$ as superscripts of quantities defined for the $\nmm$ classical particles. 
For instance, the classical mutual interaction potential, which is defined for a given number of classical particles $\nmm$, 
is denoted as $\boldwmm^\nmm$ and its explicit expression is 
\begin{equation}
 \boldwmm^\nmm(Q_1,\hdots,Q_{\nmm}) = \sum_{i=1}^{\nmm}\sum_{i > j}^{\nmm} w_{\mm}(Q_i,Q_j).
\end{equation}
Now if we consider a quantity $A^{\nmm} $ defined in the canonical 
ensemble for a given number of particles $\nmm$, its grand canonical generalization, $A^{\text{GC}}$, 
is now defined as the direct sum of $A^{\nmm} $ for all possible numbers of particles $\nmm$ 
\begin{equation}\label{eq:def_agc}
 A^{\gc}  = \bigoplus_{\nmm=0}^{\infty} A^{\nmm}. 
\end{equation}
\gj{}
Similarly, we can define the grand canonical trace operator $\text{Tr}^\gc$ acting on a grand canonical quantity $A^{\gc}$ as 
\begin{equation}
 \text{Tr}^\gc\{ A^\gc\} = \sum_{\nmm=0}^\infty \text{Tr}\{ A^{\nmm} \},
\end{equation}
where $\text{Tr}$ is the canonical trace over the degrees of freedom corresponding to $\nmm$ particles. 

As an illustration, the grand canonical extension of the classical mutual interaction is $\boldwmm^\gc$ defined as
\begin{equation}
 \boldwmm^\gc =  \bigoplus_{\nmm=0}^{\infty}  \boldwmm^\nmm(Q_1,\hdots,Q_{\nmm}) \label{eq:W^GC}.
\end{equation}
\gj{Once again, the meaning of Eq.~\ref{eq:W^GC} is that, for a particular point in the phase space in the grand canonical ensemble, the mutual interaction potential  identifies with its expression in the canonical ensemble which has the same number of particles.}
An important point to emphasis is that while $ \boldwmm^\nmm(Q_1,\hdots,Q_{\nmm})$ is simply a function with $\nmm$ arguments defined on the phase space corresponding 
to $\nmm$ particles, $ \boldwmm^\gc $ is \gj{a piecewise defined} function \textit{i.e} with a non-fixed number of arguments, as meant by the direct sum construction of  Eq. \eqref{eq:def_agc}. 

With the definition of Eq. \eqref{eq:def_agc}, we can naturally define the product and sum of two grand canonical quantities $A^{\gc}$ 
and $B^{\text{GC}}$ as follows 
\begin{equation}\label{eq:def_ab_gc}
 \begin{aligned}
 & A^{\gc} + B^{\gc} = \bigoplus_{\nmm=0}^{\infty} \big(A^{\nmm} + B^{\nmm} \big),\\
 & A^{\gc} \times  B^{\gc} = \bigoplus_{\nmm=0}^{\infty} \big(A^{\nmm}   B^{\nmm} \big),
 \end{aligned}
\end{equation}
together with the concept of evaluating a function $f(x)$ of the quantity $A^{\gc}$ 
\begin{equation}\label{eq:u_gc}
 f(A^{\gc}) = \bigoplus_{\nmm=0}^{\infty} f(A^{\nmm}). 
\end{equation}
The expressions of Eq. \eqref{eq:def_ab_gc} and Eq. \eqref{eq:u_gc} simply means that one can only make operations between objects with the same number of arguments.
This is the classical equivalent of block diagonal QM operators in the Fock space which are defined as the direct sum of all particle number sectors. 

As a pictorial example, we give here the expression of the partition function and equilibrium probability in the grand canonical ensemble $(\mu,V,T)$ of a classical system. 
For a given number of classical particle $\nmm$, this system is described by a Hamiltonian $\hmm^\nmm(\boldqb ,\boldpb )$. 
We can then define the Hamiltonian in the grand canonical ensemble, 
\begin{equation}
 \hmm^\gc = \bigoplus_{\nmm=0}^{\infty} \hmm^\nmm(\boldqb ,\boldpb ),
\end{equation}
together with the grand canonical number of particles $\nmm^\gc$
\begin{equation}
 N^\gc_\mm = \bigoplus_{\nmm=0}^{\infty} \nmm.
\end{equation}
\gj{Note that this is not the average number of particles in the grand canonical ensemble, but rather a function that returns the number of particles associated to a point in the GC phase space.}
The equilibrium density is then defined as 
\begin{equation}
  \begin{aligned}
 \eqrdmmm^\gc &= \frac{e^{-\beta (\hmm^\gc - \mu N^\gc_\mm)}}{\Xi} \\
              &= \bigoplus_{\nmm=0}^{\infty} \frac{e^{-\beta (\hmm^{\nmm} - \mu\nmm)}}{\Xi},
  \end{aligned}
\end{equation}
where the partition function $\Xi$ is defined as 
\begin{equation}  
   \begin{aligned}
    \Xi &= \text{Tr}^\gc  \{ e^{-\beta (\hmm^\gc - \mu\nmm^\gc)}\} \\
        &= \sum_{\nmm=0}^{\infty} \text{Tr}\{ e^{-\beta (\hmm^\nmm - \mu\nmm)}\} \\
        &= \sum_{\nmm=0}^{\infty} \frac{1}{N!h^{3\nmm}}\iint e^{-\beta (\hmm^\nmm - \mu\nmm)}d\boldqb d\boldpb \,\;.
   \end{aligned}
\end{equation}

\subsection{Exact Equilibrium Matrix Density \label{subsec:EDM_GC}}
In the semi-grand-canonical ensemble, the total system is divided into a set of identical classical particles, whose number $\nmm$ is allowed to vary, and a fixed number of $\nqm$ identical quantum particles, 
described by their position variables $\boldqs = \{\varqs{1}, \hdots, \varqs{\nqm}\}$.

For a given number of classical particles $\nmm$, the total Hamiltonian $\fulh$ of the system is written as 
\begin{equation}
 \label{eq:fulh_SGC}
 \fulh^{\nmm}(\boldqb ,\boldpb ,\boldqs) = \hathqm(\boldqs) + \hmm(\boldqb ,\boldpb ) + \hatboldwqmmm(\boldqb ,\boldqs),
\end{equation}
with
\begin{equation}
\label{eq:hqm}
 \hathqm = \hatboldts + \hatboldvextqm + \hatboldwqm,
\end{equation}
\begin{equation}
 \hmm = \boldtb + \boldvextmm + \boldwmm, 
\end{equation}
\begin{equation}
 \hatboldwqmmm  = \sum_{i=1}^{\nqm} \sum_{j=1}^{\nmm} \wqmmm(\varqs{i},\varqb{j}).
\end{equation}

\manu{In the semi grand canonical ensemble,} the ED matrix $\hat{\mathcal{M}}^\GC$ is expressed as
\begin{equation} \label{eq:EDM_GC1}
    \hat{\mathcal{M}}^\GC = \bigoplus_{\nmm=0}^{\infty} \dfrac{ e^{-\beta(\hathqm + \hatboldwqmmm(\boldqb))} \, e^{-\beta (\hmm(\boldqb,\boldpb ) - \mu \nmm )}}{\Xi},
\end{equation}
and  the partition function can be written as the norm of the \gj{numerator of}  Eq. \eqref{eq:EDM_GC1}, \textit{i.e.}
\begin{equation} \label{eq:Xi_2}
\Xi = \sum_{\nmm=0}^{\infty} \text{Tr}\{  e^{-\beta(\hathqm + \hatboldwqmmm(\boldqb))} \, e^{-\beta (\hmm(\boldqb,\boldpb ) - \mu \nmm )}\}. 
\end{equation}
\manu{Therefore, according to Eq. \eqref{eq:EDM_GC1}, the ED matrix is expressed through a direct sum over all possible number of classical particles.} 
From Eq. \eqref{eq:EDM_GC1}  we can see that, for a given sector $\nmm$, the EDM has a product structure of a classical probability distribution mutiplied by a QM density matrix parametrized by the $\nmm$ classical variables, as in the case of the canonicial ensemble. 
Aslo, while the QM part has precisely $\nqm$ variables, the classical part does not have a well defined number of arguments as a result of the fluctuation of the number of classical particles.

Once defined the equilibrium density matrix and partition function $\Xi$,
the corresponding grand potential $\Omega_0$ of the system is then naturally related to $\Xi$ through the usual expression
\begin{equation} \label{eq:lnXI}
\Omega_0 = - \beta ^{-1} \ln \Xi.
\end{equation}

We can establish a useful link between the semi GC ensemble and the canonical ensemble in order to reuse the previous derivations. 
The partition function can be written as 
\begin{equation} \label{eq:Xi_2}
\Xi = \sum_{\nmm=0}^{\infty} Z^\text{tot}_{\nmm} \text{e}^{\beta \mu \nmm},
\end{equation}
where  $Z^\text{tot}_{\nmm}$ is the partition function of the QM/MM system 
in the canonical ensemble for a given number $\nmm$ of classical particles. 
We recall that the latter is expressed as the trace over all QM and MM degrees of freedom, \textit{i.e.}
\begin{equation}
 \label{eq:eq_qmmm_z_GC2}
  Z^\text{tot}_{\nmm} = \intdqb \eqzqm(\boldqb)e^{-\beta\hmm(\boldqb,\boldpb )},
\end{equation}
\begin{equation}
 \label{eq:eq_qm_z_GC}
  \eqzqm(\boldqb) = \intdqs  
    \elemm{\boldqs}{e^{-\beta(\hathqm + \hatboldwqmmm(\boldqb))}}{\boldqs}.
\end{equation}
\manu{Furthermore, the EDM of Eq. \eqref{eq:EDM_GC1} in the semi GC ensemble can be written as a weighted direct sum of the EDM in canonical ensembles,} 
\begin{equation} \label{eq:EDM_GC2}
\hat{\mathcal{M}}^{\GC}  = \bigoplus_{\nmm=0}^{\infty} g_\text{eq} (\nmm) \, \hat{\mathcal{M}}(\boldqb, \boldpb ), 
\end{equation}
where $\hat{\mathcal{M}}(\boldqb, \boldpb )$ is the EDM corresponding to the QM/MM system in the canonical ensemble with $\nmm$ classical particles (therefore normalized to one), and $g_{\text{eq}}(\nmm)$ is the equilibrium probability of having the system with $\nmm$ classical particles, defined as
\begin{equation} \label{eq:geq}
g_\text{eq} (\nmm) = \frac{Z^\text{tot}_{\nmm} \text{e}^{\beta \mu \nmm}}{\Xi},
\end{equation}
which fulfils the conditions to be a probability measure 
\begin{equation} \label{eq:geq}
g_\text{eq} (\nmm)>0,\quad\sum_{\nmm=0}^{\infty}g_\text{eq} (\nmm) = 1.
\end{equation}
Another interesting property is the real-space representation of $\hat{\mathcal{M}}^{\GC}$, \textit{i.e.}
\begin{equation}\label{eq:EDM_GC3}
 \mathcal{M}^{\GC}(\boldqs, \boldqps ) = \elemm{\boldqs}{\hat{\mathcal{M}}^{\GC} }{\boldqps},
\end{equation}
which is the equilibrium density matrix of the $\nqm$ QM particles averaged over all possible number of classical particles $\nmm$. This quantity can be rewritten as a direct sum of canonical EDM and classical probabilities, 
weighted by $g_\text{eq} (\nmm)$, \textit{i.e.}
\begin{equation} \label{eq:EDM_GC4}
    \mathcal{M}^\GC (\boldqs, \boldqps ) = \bigoplus_{\nmm=0}^{\infty} g_\text{eq} (\nmm) \, \eqrdmmm_\text{eq} (\boldqb, \boldpb ) \, \eqrdmqm_\text{eq}(\boldqs, \boldqps, \boldqb),
\end{equation}
with 
\begin{equation}
 \label{eq:eq_rdmqm_GC}
 \eqrdmqm_\text{eq}(\boldqs,\boldqps,\boldqb) =\elemm{\boldqs}{\hateqrdmqm_\text{eq}(\boldqb)}{\boldqps},
\end{equation}
\begin{equation}
 \label{eq:eq_rdmqm_GC2}
  \hateqrdmqm_\text{eq}(\boldqb) = \frac{1}{\eqzqm(\boldqb)}e^{-\beta(\hathqm + \hatboldwqmmm(\boldqb))},
\end{equation}
\begin{equation}
 \label{eq:eq_rdmmm_GC}
 \eqrdmmm_\text{eq} (\boldpb ,\boldqb) = \frac{\eqzqm(\boldqb)}{Z^\text{tot}_{\nmm}}e^{-\beta\hmm(\boldqb,\boldpb )}.
\end{equation}


\subsection{Exact formulation of the grand potential for a
QM/MM system within a variational framework \label{subsec:FormVar}}

\gj{With the definitions of Sec. \ref{subsec:EDM_GC}, we can now turn to our objective which is to}
reformulate the grand potential $\Omega_0$ as a variational problem over QM and MM quantities, \gj{following a route similar to what is}  done with the Helmholtz free energy \gj{in the main paper}. 

The first step is to notice that the form of the EDM $\hat{\mathcal{M}}^{\GC}$ of Eq. \eqref{eq:EDM_GC4} is a special case of more general semi grand canonical density matrices of the form 
\begin{equation} \label{eq:DMGC1}
\dmgc = \bigoplus_{\nmm=0}^{\infty} g (\nmm) \, \eqrdmmm_{\nmm} (\boldqb, \boldpb ) \, \hateqrdmqm_{\nmm} (\boldqb),
\end{equation}
where now $g (\nmm)$ is a generic classical probability distribution to find the system with a given number of classical particles, 
$\eqrdmmm_{\nmm} (\boldqb, \boldpb )$ is a generic classical probability distributions for the variables $(\boldqb, \boldpb )$, and $\hateqrdmqm_{\nmm} (\boldqb)$ is a generic QM density matrix parametrized by  the degrees of freedom of the $\nmm$ classical particles.

Therefore, as previously done in the canonical ensemble, we can  express the grand potential as a minimization over all possible semi GC density matrices of the form of Eq. \eqref{eq:DMGC1}
\begin{equation} \label{eq:Var_prin_GC1}
\Omega_0 = \min_{\dmgc} \Omega[ \dmgc ],    
\end{equation}
where the grand potential functional $\Omega[ \dmgc ]$ is given by  
\begin{equation}\label{eq:def_omega}
  \Omega[ \dmgc ] = \ecano[\dmgc] - \mu N[\dmgc],
\end{equation}
where the average number of classical particles is simply expressed as 
\begin{equation}
N[\dmgc] = \sum_{\nmm=0}^{\infty}g(\nmm)\,\, \nmm ,
\end{equation}
and the average "free energy" is given by
\begin{equation}
 \label{eq:pot_cano}
 \begin{aligned}
  \ecano[\dmgc] = \uint[\dmgc] - \kb T \entrop[\dmgc].
 \end{aligned}
\end{equation}
In Eq. \eqref{eq:pot_cano}, $\ecano[\dmgc]$ is decomposed in terms of an internal energy functional $\uint[\dmgc]$ 
and a dimensionless entropy functional $\entrop[\dmgc]$, which can be expressed as averages of quantities 
over all possible number of classical particles. 
More precisely, $\uint[\dmgc]$ is defined as 
\begin{equation}
 \uint[\dmgc] = \sum_{\nmm=0}^{\infty} g(\nmm) E[\hat{\eqrdmqm}_{\nmm}, \eqrdmmm_{\nmm}],
\end{equation}
where $E[\hat{\eqrdmqm}_{\nmm}, \eqrdmmm_{\nmm}]$ is the internal energy functional associated to the couple 
$(\hat{\eqrdmqm}_{\nmm}, \eqrdmmm_{\nmm})$ defined as 
\begin{equation}
E[\hat{\eqrdmqm}_{\nmm}, \eqrdmmm_{\nmm}] = \tr\big\{\hat{\eqrdmqm}_{\nmm} \eqrdmmm_{\nmm}\fulh^\nmm\big\},
\end{equation}
and similarly the dimensionless entropy functional is obtained as 
\begin{equation}\label{eq:entrop_clean}
 \entrop[\dmgc] = -\sum_{\nmm=0}^{\infty}g(\nmm)\tr\big\{\hat{\eqrdmqm}_{\nmm} \eqrdmmm_{\nmm}
 \log\big(g(\nmm)\hat{\eqrdmqm}_{\nmm} \eqrdmmm_{\nmm}\big)\big\}.
\end{equation}

In order to \gj{prove} that the present definitions of the grand potential functional (\textit{i.e.} Eqs. \eqref{eq:def_omega}-\eqref{eq:entrop_clean}) satisfy a variational formulation of the grand potential $\Omega_0$, we \gj{resort} the same methodology as in  the canonical ensemble, \gj{that is to} use Jensen's inequality. 
To be able to use \gj{this convex analysis tool}, we first need to rewrite the grand potential as an average over a probability measure. 
We start with the expanded expression of the grand potential functional
\begin{equation} \label{eq:Omega_func}
\Omega[\dmgc] = \sum_{\nmm=0}^{\infty} g(\nmm) \text{Tr}\big\{\hat{\eqrdmqm}_{\nmm} \eqrdmmm_{\nmm} \big[\fulh  -\mu \nmm
+ \beta^{-1} \log \big(g(\nmm)\hat{\eqrdmqm}_{\nmm} \eqrdmmm_{\nmm}\big)\big].
\end{equation}
Since $\dmgc$ is a density matrix, it can then be mapped to a probability measure $m({\bf X})$, where ${\bf X}$ represents all classical and quantum degrees of freedom
\begin{equation}
 \dmgc \equiv m({\bf x}),\quad m({\bf x}) >0, \quad \int d{\bf X} m({\bf X})=1,
\end{equation}
the symbol $\int d{\bf X}$ representing the summation over all degrees of freedom, \textit{i.e.}
\begin{equation}
\int d{\bf X}  \equiv \sum_{\nmm=0}^{\infty} \text{Tr} = \sum_{\nmm=0}^{\infty}\int d\boldqb d\boldpb  \tr_{\qm}.
\end{equation}
Therefore, the functional $\Omega$ of Eq. \eqref{eq:Omega_func} can be formally rewritten as 
\begin{equation}
  \Omega[m] = \int d{\bf X} m({\bf X})\big(\epsilon({\bf X})+\kb T \log(m({\bf X})),
\end{equation}
where 
\begin{equation}
\epsilon({\bf X}) \equiv \fulh  -\mu \nmm.
\end{equation}

As a consequence, similarly to what have been done previously in the case of the canonical ensemble, 
we can then rewrite $\Omega[m]$ as 
\begin{equation}
\Omega[m] = -\kb T\langle \log(\frac{e^{-\beta\epsilon}}{m}) \rangle_{m},
\end{equation}
where now it appears as strictly convex with respect to the probability measure $m$, and then use Jensen's inequality to prove that it admits a unique minimum corresponding to the grand potential of Eq. \eqref{eq:lnXI}, and the only minimizer is the EDM of Eq. \eqref{eq:EDM_GC4}.

\subsection{Levy-Lieb constrained-search formulation of the grand potential \label{subsec:LevyLieb}}
Now that the variational formulation has been established, using the Levy-Lieb constrained search, we can then rewrite it as a functional minimization over  the one-body densities, both classical and quantum.

We begin by splitting the functional $\Omega[\dmgc]$ in terms of an intrinsic part and an external part
\begin{equation} \label{eq:Omega=int+ext}
    \Omega [\dmgc] = \omega_\text{int}[\dmgc] + \omega_\text{ext} [\dmgc],
\end{equation}
where
\begin{equation} \label{eq:Omega_int}
    \omega_\text{int}[\dmgc] = \sum_{\nmm=0}^{\infty} g(\nmm) \left[  E_\qm [\hat{\eqrdmqm}_{\nmm}, \eqrdmmm_{\nmm} ] + E_\mm [\eqrdmmm_{\nmm}] + E_\qm^\mm [\hat{\eqrdmqm}_{\nmm},\eqrdmmm_{\nmm}] - \mu \nmm + \beta^{-1} \log \big(g(\nmm)\hat{\eqrdmqm}_{\nmm}\eqrdmmm_{\nmm}\big)  \right],
\end{equation} 
\begin{equation}\label{eq:Omega_ext}
    \omega_\text{ext} [\dmgc] = \sum_{\nmm=0}^{\infty} g(\nmm) \left[  E_\qm^\text{ext} [\hat{\eqrdmqm}_{\nmm},\eqrdmmm_{\nmm}] + E_\mm^\text{ext} [\hat{\eqrdmqm}_{\nmm},\eqrdmmm_{\nmm}]  \right]. 
\end{equation}
The functional $\omega_\text{int}[\dmgc]$ of Eq. \eqref{eq:Omega_int} gathers all intrinsic properties of the system, which contains the QM and MM internal energies, the QM/MM interaction together with the total entropy of the system in the semi grand canonical ensemble. On the other hand, $\omega_\text{ext} [\dmgc]$ of Eq.  \eqref{eq:Omega_ext} gathers the interaction of both the QM and MM region with the external potentials. 

Since we assumed that the external potentials are  sums of one-body external potentials, $E_\qm^\text{ext}$ and $E_\mm^\text{ext}$ can be computed with only the one-body densities, both quantum and classical. 
More precisely, we introduce the semi grand canonical classical density 
\begin{equation}
\densmm^{\GC}(Q) = \sum_{\nmm=0}^{\infty} g(\nmm) \int dQ_2 \hdots dQ_{\nmm} \eqrdmmm_{\nmm}(Q,Q_2,\hdots,Q_{\nmm}),
\end{equation}
and its quantum counter part 
\begin{equation}\label{eq:rhoelGC}
\densqm^{\GC} (q) = \sum_{\nmm=0}^{\infty} g(\nmm) \int d\boldqb\eqrdmmm_{\nmm}(\boldqb) \int dq_2\hdots dq_{\nqm} {\eqrdmqm}_{\nmm}(q,q_2,\hdots,q_{\nqm},\boldqb),
\end{equation}
where 
\begin{equation}
 {\eqrdmqm}_{\nmm}(q_1,q_2,\hdots,q_{\nqm},\boldqb) = \elemm{q_1,\hdots,q_\nqm}{\hat{\eqrdmqm}_{\nmm}(\boldqb)}{q_1,\hdots,q_\nqm}.
\end{equation}
With these definitions, the external part of the grand potential $\omega_\text{ext}$, in Eq.~\eqref{eq:Omega_ext}, becomes
\begin{equation}
    \omega_\text{ext} [\densqm^{\GC}, \densmm^{\GC}] = \left( \vextqm \mid \densqm^{\GC} \right) + \left( v_{\text{ext}}^{{\mm}} \mid \densmm^{\GC} \right),
\end{equation}
with
\begin{equation}
    \left( v_{\text{ext}}^{{\mm}} \mid \densmm^{\GC} \right) \equiv \int dQ \densmm^{\GC} (Q ) \, v_{\text{ext}}^{{\mm}}(Q) \, ,
\end{equation}
\begin{equation}
    \left( \vextqm \mid \densqm^{\GC} \right) \equiv \int \densqm^{\GC}(q) \, \vextqm(q).
\end{equation}

We can then rewrite the minimization of Eq. \eqref{eq:Var_prin_GC1} using the Levy-Lieb constrained search  
\begin{equation}
\Omega_0 = \min_{(\densqm^{\GC},\densmm^{\GC})}\quad\min_{\dmgc \rightarrow(\densqm^{\GC},\densmm^{\GC}) } \Omega[ \dmgc ],    
\end{equation}
and by introducing the Levy-Lieb QM/MM universal functional $\Omega_\text{int}[\densqm^{\GC}, \densmm^{\GC}]$ in the semi-grand-canonical ensemble as 
\begin{equation} \label{eq:Omega_LL1}
    \Omega_\text{int}[\densqm^{\GC}, \densmm^{\GC}] = \min_{\dmgc \rightarrow (\densqm^{\GC}, \densmm^{\GC})} \omega_\text{int}[\dmgc],
\end{equation}
we can then rewrite the grand canonical potential as a minimization over only the one-body QM and MM densities 
\begin{equation}
\Omega_0 =  \min_{(\densqm^{\GC},\densmm^{\GC})} \big\{ \Omega_\text{int}[\densqm^{\GC}, \densmm^{\GC}] + 
\left( \vextqm \mid \densqm^{\GC} \right) + \left( v_{\text{ext}}^{{\mm}}\mid \densmm^{\GC} \right) \big\}. 
\end{equation}
Although the exact expression of this functional is unknown, as in the case of the canonical ensemble, we choose to express it using known universal functional for the purely quantum and classical components. Based on Eq.~\eqref{eq:Omega_int}, we obtain
\begin{equation} \label{eq:Omega_LL2}
    \Omega_\text{int}[\densqm^{\GC}, \densmm^{\GC} ] = \fqm[\densqm^{\GC}] + \Omega_\mm [\densmm^{\GC}] + \Omega_{\qm}^\mm[\densqm^{\GC}, \densmm^{\GC}],
\end{equation}
where $\fqm[\densqm^{\GC}]$ is the universal purely quantum
one-body density functional,
\begin{equation}  \label{eq:mermin}
 \fqm[\densqm^{\GC}] = \min_{\hatrdmqm\rightarrow\densqm^{\GC} } 
 \text{Tr}\{\hatrdmqm\big[ \hatboldts  + \hatboldwqm + \kb T  \log(\hatrdmqm)\big]\},
\end{equation}
and $\Omega_\mm[\densmm^{\GC}] $ is the universal classical one-body functional of Evans\cite{Evans-AP-79}, 
but expressed using the constrained search formalism of Levy-Lieb. 
Lastly, in Eq.~\eqref{eq:Omega_LL2}, we define the term $\Omega_\qm^\mm$, which represents the interactions between the QM and the MM parts
\begin{equation}
    \Omega_{\qm}^\mm [\densqm^{\GC}, \densmm^{\GC}] = \Omega_\text{int}[\densqm^{\GC}, \densmm^{\GC}] - \fqm [\densqm^{\GC}] - \Omega_\mm[\densmm^{\GC}].
\end{equation}
In the case of a two-body QM/MM interaction, 
we can then decompose this complex term into mean-field and correlation parts, namely
\begin{equation} \label{eq:Omega_qmmm[rhomel,rhomm]}
    \Omega_{\qm}^\mm [\densqm^{\GC}, \densmm^{\GC}] = \varepsilon_{\qm}^\mm [\densqm^{\GC}, \densmm^{\GC}] + \delta \Omega_{\qm}^\mm [\densqm^{\GC}, \densmm^{\GC}],
\end{equation}
with
\begin{equation} \label{eq:eps_qmmmGC[rhomel,rhomm]}
    \varepsilon_{\qm}^\mm [\densqm^{\GC}, \densmm^{\GC}] = \iint \densmm^{\GC}(Q ) \, v_\qm^\mm(Q,q) \, \densqm^{\GC}(q) \, \mathrm{d} Q  \, dq,
\end{equation}
\begin{equation} \label{eq:dOmega_qmmm[rhomel,rhomm]}
    \delta \Omega_{\qm}^\mm [\densqm^{\GC}, \densmm^{\GC}] = \Omega_\text{int}[\densqm^{\GC}, \densmm^{\GC}] - \fqm[\densqm^{\GC}] - E_\mm[\densmm^{\GC}] - \varepsilon_{\qm}^\mm [\densqm^{\GC}, \densmm^{\GC}],
\end{equation}
where $\varepsilon_{\qm}^\mm [\densqm^{\GC}, \densmm^{\GC}]$ is the mean-field interaction term directly computable, while $\delta \Omega_{\qm}^\mm [\densqm^{\GC}, \densmm^{\GC}]$ is the unknown term containing QM/MM correlation. 
We emphasize that Eq.~\eqref{eq:Omega_LL1} constitutes, to our knowledge, the first exact formulation of a Levy-Lieb universal functional for a QM/MM system in the semi-grand-canonical ensemble.

{Eventually}, the grand potential $\Omega_0$ is expressed as minimizations of a functional with respect to the one-body densities $\densqm$ and $\densmm^{\GC}$, 
\begin{equation} \label{eq:Omega_final}
    \Omega = \min_{\densqm^{\GC}, \densmm^{\GC}} \big\{ \fqm[\densqm^{\GC}] + \Omega_\mm[\densmm^{\GC}] + \delta \Omega_\qm^\mm[\densqm^{\GC}, \densmm^{\GC}]  + \varepsilon_{\qm}^\mm [\densqm^{\GC}, \densmm^{\GC}] + (v^{\text{ext}}_{\qm} | \densqm^{\GC}) + (v^{\text{ext}}_\mm | \densmm^{\GC}) \big\}.
\end{equation}
Although  Eq.~\eqref{eq:Omega_final} is exact, the correlation term (\textit{i.e.}~$\delta \Omega_\qm^\mm$) is in practice incalculable.